\documentclass[journal]{IEEEtran}

\usepackage[utf8]{inputenc} 
\usepackage[normalem]{ulem}
\usepackage[T1]{fontenc}
\usepackage{url}
\usepackage{ifthen}
\usepackage{multirow}

\usepackage{graphicx,color,epsfig,rotating,subfigure}       
\usepackage{algorithm,algcompatible,xcolor,cite}
\usepackage{graphicx,color,epsfig,rotating,subfigure}
\usepackage{amsfonts,amsmath,amssymb, bm}
\usepackage{algorithm}
\usepackage{subfigure}
\usepackage{algpseudocode}
 \newcommand{\um}{\textcolor{blue}}  
 
 \newcommand{\pn}{\textcolor{teal}}  
 \newcommand{\jm}{\textcolor{red}} 

\long\def\comment#1{}

\newfont{\bbb}{msbm10 scaled 700}

\newfont{\bb}{msbm10 scaled 1100}

\newcommand{\RR}{\mbox{\bb R}}

\newcommand{\EE}{\mbox{\bb E}}

\renewcommand{\arg}{{\hbox{arg}}}

\newcommand{\SNR}{{\sf SNR}}

\newcommand{\trasp}{{\sf T}}
\newcommand{\transp}{{\sf T}}

\usepackage{graphicx}
\usepackage{tabularx,booktabs,hhline}
\usepackage{color, colortbl}
\definecolor{LightCyan}{rgb}{0.88,1,1}
\definecolor{lightgray}{gray}{0.95}
\usepackage{multirow}
\usepackage{soul} 
\usepackage{amsthm}
\newtheorem{assumption}{Assumption}

\def \mM \mathbf{M}
\def \mQ \mathbf{Q}
\def \mS \mathbf{S}
\def \mV \mathbf{V}
\def \mU \mathbf{U}
\def \mX \mathbf{X}
\begin{document}

\title{Matrix Approximation with Side Information:\\ When Column Sampling is Enough} 

\author{Jeongmin Chae$^{\dagger}$, Praneeth Narayanamurthy$^{\dagger}$, Selin Bac$^{*}$, 
Shaama Mallikarjun Sharada$^{*}$ and Urbashi Mitra$^{\dagger}$
\thanks{$^{\dagger}$ J. Chae, P. Narayanamurthy and U.Mitra are with the Department of Electrical and Computer Engineering, University of Southern California, Los Angeles, USA. E-mail: \{chaej,praneeth, ubli\}@usc.edu.}
\thanks{
$^{*}$ S. Bac and S. Mallikarjun Sharada are with the Department of Chemical Engineering and Materials Science, University of Southern California, Los Angeles, USA. E-mail: \{bacbilgi, ssharada\}@usc.edu.}}

\newtheorem{theorem}{Theorem}
\newtheorem{definition}{Definition}
\newtheorem{lemma}{Lemma}
\newtheorem{corollary}{Corollary}
\newtheorem{remark}{Remark}
\newtheorem{proposition}{Proposition}
\newcommand{\argmin}{\operatornamewithlimits{argmin}}
\newcommand{\argmax}{\operatornamewithlimits{argmax}}

\renewcommand{\subsubsection}[1]{\noindent {\bf #1. }}
\renewcommand{\qedsymbol}{$\boxtimes$}

\newcommand{\vqs}{\mathbf{V}_{QS}}
\newcommand{\vqsh}{\hat{\mathbf{V}}_{QS}}
\newcommand{\uqs}{\mathbf{U}_{QS}}
\newcommand{\sqs}{\mathbf{\Sigma}_{QS}}
\newcommand{\uM}{\mathbf{U}_{M}}
\newcommand{\sM}{\mathbf{\Sigma}_{M}}
\newcommand{\vM}{\mathbf{V}_{M}}
\newcommand{\vMh}{\hat{\mathbf{V}}_{M}}
\newcommand{\uMb}{\mathbf{U}_{M,\perp}}
\newcommand{\sMb}{\mathbf{\Sigma}_{M,\perp}}
\newcommand{\vMb}{\mathbf{V}_{M,\perp}}
\newcommand{\vqsb}{\mathbf{V}_{QS,\perp}}
\newcommand{\uqsb}{\mathbf{U}_{QS,\perp}}
\newcommand{\sqsb}{\mathbf{\Sigma}_{QS,\perp}}
\newcommand{\uA}{\mathbf{U}_{A}}

\maketitle

\begin{abstract}
A novel matrix approximation problem is considered herein:
observations based on a few fully sampled columns and quasi-polynomial structural side information are exploited. The framework
is motivated by quantum chemistry problems wherein full
matrix computation is expensive, and partial computations
only lead to column information. The proposed algorithm
successfully estimates the column and row space of a true matrix given
a priori structural knowledge of the true matrix. A theoretical spectral error
bound is provided, which captures the possible inaccuracies
of the side information. { The error bound proves it scales in its signal-to-noise ($\mbox{SNR}$) ratio as $\mbox{SNR}^{-1}$}. The proposed algorithm is validated via simulations which enable the characterization of
the amount of information provided by the quasi-polynomial
side information.
\end{abstract}


\section{Introduction}\label{sec:intro}

\begin{figure}[t!]
    \centering
\includegraphics[width=8.5cm]{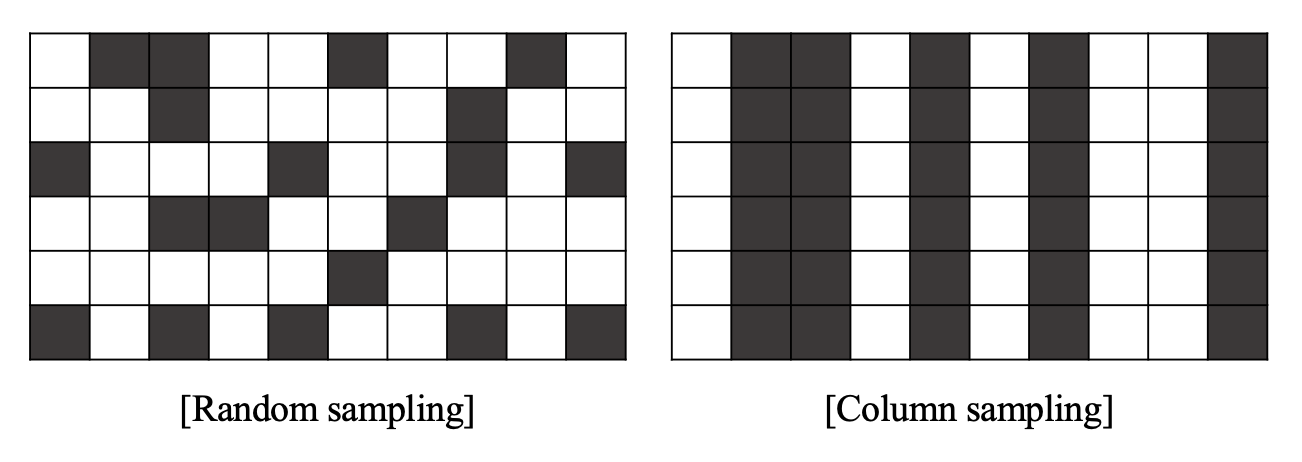}
    \caption{An example of random sampling and column-based sampling when a matrix $\mathbf{M} \in \RR^{6 \times 10}$. The black entries indicate observed samples.}
   \label{fig:sampling}
\end{figure}

Matrix completion wherein missing components of a matrix are imputed by exploiting key structural information has found wide application over the years \cite{elmahdy2020matrix,chiang2015matrix,zhang2019inductive,xu2013speedup}.  Conventional matrix completion assumes that sampled entries are collected uniformly randomly \cite{candes_exact_2009,candes2010matrix}, however, recent work suggests that many applications (e.g. computer vision, bioinformatics, economics, and data science) do not admit a uniform sampling strategy 
\cite{cai2016structured,farias2021learning,liu2017new}. 
Herein, we consider a problem motivated by quantum chemistry \cite{quiton_matrix_2020,bac2022matrix,quiton2022toward} where the full matrix can be computed, albeit via very expensive computations; however partial computations can be done, but only for columns of the true matrix (this sampling strategy is depicted in Figure~\ref{fig:sampling}).  

\newcommand{\polylog}{\mathrm{polylog}}
In \cite{candes_exact_2009,candes2010matrix},  the theoretical guarantee that $O(nr \mu \polylog n)$ observations (chosen uniformly at random) are sufficient for exact recovery for a $n \times n$ matrix of rank $r$ under the assumptions that the coherences $\mu$ of row and column subspace are bounded by some small positive value. Integral is the assumption that both the column and row space of the true matrix must be incoherent.

{ While there are several studies that reduce the sample complexity when access to side information is provided \cite{xu2013speedup,natarajan2014inductive,chiang2015matrix} these works still impose the assumption of uniform random sampling and access to \textit{perfect} side information\footnote{Imperfect side information is considered in \cite{chiang2015matrix} with uniform sampling. }. On the other hand, there is  another line of work that addresses non-uniform sampling \cite{foucart2020weighted,negahban2012restricted,liu2017new,xu2015cur} but none of these allow for full column sampling that we consider in this paper. Furthermore, in all these works, it is assumed that the underlying matrix is low-rank unlike our target application wherein the goal is to provide a low-rank approximation for a high-rank matrix. }

Herein, we focus on matrix approximation versus matrix recovery. Thus, our goal is to construct a low rank-$r$ approximation, $\hat{\mathbf{M}}$, of the true matrix $\mathbf{M}$ which has rank $k$ where $r \leq k$. Two popular formulations that analyze matrix approximation include approximation from sketches \cite{matrix_recovery_sketch, matrix_recovery_linear}, and the CUR decomposition \cite{drineas2008relative,mahoney2009cur}. The former captures dense, and global measurements of the matrix (in the form of random linear projections of the complete matrix), while the latter captures sparse, and local measurements (in the form of observing a small subset of rows and columns of the original matrix)\footnote{Matrix completion is the most sparse, and highly local information analogue.}. We {consider} the CUR decomposition approach since it is most applicable to our problem setting. While there exists a plethora of approaches that develop sophisticated sampling schemes to perform CUR decomposition, there is very little work that deals with missing data. A recent paper, CUR+ \cite{xu2015cur} remedies this and computes an error bound with sample complexity $O(n r \ln r)$, but it still employs fully sampled rows and columns as well as additional random samples of the original matrix. Thus, although CUR+ offers an important benchmark to our method, it cannot be applied directly to our problem where row sampling cannot occur. We provide a detailed comparison in Sec. \ref{sec:disc}.


 The contributions of this work are as follows:

\begin{enumerate}
\item The novel problem framework is delineated.  
We propose a matrix approximation method based on randomly, but fully sampled columns coupled with side information that is captured via a quasi-polynomial structure.  The side information is assumed imperfect. We further assume that our true matrix is high-rank, but seek a low-rank approximation. 
\item The quasi-polynomial matrix approximation (QPMA) algorithm is introduced based on a well-designed objective function and an associated approximation strategy.
\item { A theoretical spectral bound on the reconstruction error achieved by QPMA is derived.  This bound is shown to be only slightly worse, with respect to order of key parameters, to that achievable by prior matrix approximation strategies with a significantly weaker assumption.  In particular,  Theorem~\ref{eq:thm1}, shows that the matrix can be recovered when row space information is provided that is close to that of the true matrix. }
\item  QPMA is compared to CUR+ on synthetic data which enables the characterization of the amount of information about the row space that is afforded by the quasi-polynomial structure relative to the number of rows needed by CUR+.
\item QPMA is also assessed via application to the original quantum chemistry problem.
\end{enumerate}
While our strategy is motivated by a specific application, we believe our algorithm and analysis will have greater applicability to problems wherein side information can be succinctly captured by row space constraints. Characterizing the applicability of our methods to more general problems is an avenue for future work. In our prior work \cite{quiton2022toward}, an algorithm for column sampling coupled with quasi-polynomial side information was proposed and numerically
shown to offer good performance.  A challenge with the proposed algorithm in \cite{quiton2022toward} was control of the rank of the approximated matrix.  With the modified approach herein, we can carefully control for rank while providing theoretical guarantees that are based on algorithm and system parameters.

This paper is organized as follows. Section II introduces the quantum chemistry application that motivates this work and provides the background of the system model. After that, the formal problem setting and the description of the proposed algorithm are provided in Section III. The main result of this paper is to understand the spectral reconstruction error coupled with the key parameters such as a target rank, a true rank and a polynomial degree.  This result is presented in Section IV with the discussions on time complexity and comparison with CUR+. Further, the key simulation results to evaluate the main theorem are provided in Section V. The remainder of the paper is the proof of the theorem and key lemmas.

The following notation is adopted herein.
We define $[m]\doteq\{1,\dots,m\}$. Bold upper case letters $\mathbf{M}$ denote matrices. For a given set $\mathcal{C}$, $\vert \mathcal{C} \vert$ denotes the cardinality of the set. {{We use ${\Vert \cdot \Vert}$ to denote the spectral norm unless otherwise specified.}} We use $\overset{\text{SVD}}{=}$, $\overset{r-\text{SVD}}{=}$ to denote the singular value decomposition (SVD) and the reduced (rank-$r$) SVD of a matrix, respectively. $\sigma_{r}(\mathbf{M})$ refers to the $r$-th largest singular value of a matrix $\mathbf{M}$. {{$\mathbf{M}^{\dagger}$ denotes the pseudo-inverse of $\mathbf{M}$. Finally, throughout this paper, with a slight abuse of terminology, we use the terms column and row space of a matrix, $\mathbf{M}$ to mean the {\em best $r$-dimensional} approximation for the respective spaces. }}
 
\section{Motivating Application}
\label{sec:motiv}
We provide the motivation for this work and its applications.
In the study of chemical reactions using quantum chemistry methods, Variational Transition State Theory (VTST) is a technique
for calculating reaction rate coefficients that describe kinetics \cite{garrett1980variational}. 
VTST suffers from high computational cost as it requires the calculation of expensive quantum mechanical Hessians of energy at several points constituting the minimum energy path (MEP) of a reaction. 
Prior efforts towards reducing computational effort  include interpolated VTST (I-VTST), which fits splines under tension to  energies, gradients, and Hessians calculated at arbitrary points on the MEP \cite{gonzalez1991interpolated}.
In our prior work \cite{quiton_matrix_2020,bac2022matrix}, we showed that randomized sampling coupled with an algebraic variety constraint \cite{ongie_algebraic_2017} could
accurately complete an incomplete matrix of Hessian eigenvalues constituting the MEP when only a small, randomly sampled set of elements are available. In particular, the algebraic variety constraint is well-matched to this problem as, within the reaction path Hamiltonian (RPH) framework\cite{miller1980reaction}, the harmonic potential energy terms  $V(s,{\bf q})$ {{are formulated into {a} polynomial expression of the eigenvalues $\{w^2_{k}\}$ of Hessian matrix and displacements along vibrational normal modes $\{q^2_{k}\}$ as
\begin{align}
    V(s,{\bf q})=V_{0}(s)+\sum_{k=1}^{n} w_k^2 q_k^2,
\end{align} where $n=3n_a-7$  indicates the number of vibrational modes that are orthogonal to the reaction coordinate, $n_{a}$ is the number of atoms and $V_{0}(s)$ is the potential energy at a point $s$ on the MEP.
}}

While our algorithm proposed in \cite{quiton_matrix_2020} was computationally efficient and provided a proof-of-concept, it assumed randomized sampling, whereas, pragmatically one can compute one Hessian at a time, which corresponds to one column of the true matrix. 



The true matrix $\mathbf{M}$ of Hessian eigenvalues constituting the MEP is constructed by the potential energy term of the reaction path Hamiltonian \cite{miller1980reaction,page1988evaluating}.
The true matrix $\mathbf{M}$ is given by 
\begin{align}
    \mathbf{M} \in \RR^{n \times m}=
    \begin{bmatrix}
    \omega_1^{2}(s_1) & 
    \omega_1^{2}(s_2) & 
    \dots & \omega_1^{2}(s_m)  \\
    \omega_2^{2}(s_1) & 
    \omega_2^{2}(s_2) & 
    \dots & \omega_2^{2}(s_m)    \\
    \vdots \\
    \omega_{n}^{2}(s_1) & 
    \omega_{n}^{2}(s_2) & 
    \dots & \omega_{n}^{2}(s_m)
    \end{bmatrix}\nonumber.
\end{align}
$\{\omega_i\}, i \in [n]$,  constitutes the set of vibrational frequencies of the system obtained upon projecting out the reaction coordinate, translations, and rotations from the Hessian.
Each column $\mathbf{M}_{j}, j \in [m]$, is comprised of $n$ eigenvalues {$\{w_{i}^2\}, i \in [n]$, of the projected quantum mechanical Hessian matrix.
The reaction coordinate is parameterized by $s_i$, where $i \in [m]$, defined to be zero at the transition state, negative in the reactant region (with reactant represented by $s_1$), and positive in the product region (with product represented by $s_m$). The goal is to approximate $\mathbf{M}$ given a few full columns in a way that VTST rate coefficients can be estimated with reasonable accuracy. 

\section{Problem Formulation and Algorithm}
We next present the concrete problem formulation, the proposed optimization strategy, and finally our main guarantee. 

\subsection{Problem setting} 
Let $\mathbf{M} \in \RR^{n \times m}$ be the true matrix of rank $k$. In this paper, we consider the problem of obtaining a rank-$r$ approximation of $\mathbf{M}$ from $d$ randomly sampled columns. In particular, we consider the following regime 
\begin{align}
    \underbrace{r}_{\text{target rank}} \leq \underbrace{d}_{\text{\# of sampled columns}} \leq \underbrace{k}_{\text{true rank}} \nonumber
\end{align}

In contrast to traditional Matrix Completion, we seek a lower rank approximation of the matrix (w.r.t. the true rank). This allows us to obtain our main guarantees even when the number of columns sampled is smaller than the actual rank.


To describe the chemical rate reaction processes, as mentioned in Section II, given a list of reaction coordinate values, ${\mathbf{ s}}=[s_1,\dots,s_m]$ and polynomial order $l$ we model $\mathbf{M}$ as 
\begin{align}
    \mathbf{M} = \mathbf{Q}\mathbf{S} + \mathbf{E}\label{eq:structure_M}
\end{align}
where, $\mathbf{Q} \in \RR^{n \times l}$ is an unknown polynomial coefficient matrix, the structural side information matrix, $\mathbf{S} \in \RR^{l \times m}$ encodes the known polynomial information (described next) and $\mathbf{E}$ is the perturbation/noise matrix. 

Per Section II, the eigenvalues of the Hessian matrix $\mathbf{M}$ is quasi-polynomial, we assume that the side information $\mathbf{S}$ has the following structure
{{
\begin{align}\label{eq:S}
{\bf S} = 
    \begin{bmatrix}
    1 & 1 & \dots & 1 \\
    s_1 & s_2 & \dots & s_m \\
    s_1^2 & s_2^2 & \dots & s_m^2 \\
    \vdots & \vdots & \vdots & \vdots\\
    s_1^{l-1} & s_2^{l-1} & \dots & s_m^{l-1} \\
    \end{bmatrix}.
\end{align} }}


As mentioned previously, we observe a subset of the columns of $\mathbf{M}$. This column sampling operation  $\mathbf{\Psi}$ is defined as follows. Let $\mathcal{C}=\{c_1,\dots,c_d\}\subset [m]$ denote the set of sampled column indices. Clearly 
$\vert\mathcal{C}\vert=d$. Then, ${\mathbf{ \Psi}} \in \{0,1\}^{m \times d}$ is
\begin{align}
    {\mathbf {\Psi}}  \doteq {\mathbf{I}}_{\mathcal{C}},\nonumber
\end{align} 
where $\mathbf{I}$ is the identity matrix of dimension $m$ and the notation $\mathbf{I}_{\mathcal{C}}$ means that we consider the sub-matrix of $\mathbf{I}$ formed by its columns indexed by entries in the set $\mathcal{C}$ \footnote{For example, when $m=4$, $\mathcal{C}=\{1,3,4\}$, i.e., $d = 3$,  
\begin{align}
    {\mathbf{\Psi}} =
    \begin{bmatrix}
    1 & 0 & 0\\
    0 & 0 & 0\\
    0 & 1 & 0 \\
    0 & 0 & 1\\
    \end{bmatrix}\nonumber.
\end{align}
}. 
{Thus, the observed matrix, $\mathbf{A}$, can be equivalently expressed as
\begin{align}
    \mathbf{A}=\mathbf{M}{\mathbf{\Psi}}\nonumber.
\end{align} 
Table I summarizes the parameters for the introduced matrices.}

\begin{table}[ht!]
    \caption{Description of various matrix ranks}
    \centering
    \begin{tabular}{c|c|c}
    \toprule
     Matrix & Rank & Relationship \\  \hline
     {\bf M} &  $k$ & \multirow{4}{*}{$r \leq l,d \leq k$}\\   \cline{1-2}
     {\bf QS} &  $l$ &\\  \cline{1-2}
     ${\bf A=M\Psi}$ & $\leq d$ & \\ 
     \bottomrule
    \end{tabular}
    \label{tb:tb2}
\end{table} 

Before setting up the optimization problem, we define the following quantities. We denote the SVD of the true matrix, $\mathbf{M}$ as 
\begin{align}\label{eq:svd_M} 
     \mathbf{M}&\stackrel{\text{SVD}}{=}{\mathbf{ U}\mathbf{\Sigma} \mathbf{V}^{\trasp}} = \underbrace{\uM\sM\vM^\transp}_{\text{rank-}r\text{-approximation}} + \underbrace{\uMb\sMb\vMb^\transp}_{\text{remainder}} 
\end{align}
 
Notice that when the target rank $r$ is smaller than the true rank $k$, the second term above is non-zero. Similarly, we define the SVD of $\mathbf{QS}$ as
\begin{align}\label{eq:svd_qs}
\mathbf{QS}&\stackrel{\text{SVD}}{=}\tilde{\mathbf{U}} \tilde{\mathbf{\Sigma}} \tilde{\mathbf{V}}^{\trasp} = \underbrace{\uqs\sqs\vqs^\transp}_{\text{rank-}r\text{-approximation}} + \underbrace{\uqsb\sqsb\vqsb^\transp}_{\text{remainder}}
\end{align}

\renewcommand{\mM}{{\mathbf{M}}}

\subsection{Quasi-Polynomial Matrix Approximation Algorithm}\label{sec:4_2}
We next introduce the proposed optimization strategy, Quasi Polynomial Matrix Approximation (QPMA).  {{Note that if the column and row space information of $\mM$, i.e., $\uM$ and $\vM$ respectively,}} were known, a natural way to cast the optimization that takes into account the structural information including the desired rank $r$ approximation of $\mathbf{M}$ is as 
\begin{gather}\label{eq:final_opt}
    \;\;\;\min_{\mathbf{Z}}\;\;   \left\Vert \mathbf{A}-\uM\mathbf{Z}{\vM^{\trasp}}{\mathbf{\Psi}}\right\Vert_{F}^{2}
\end{gather} 
However, since we do not know $\uM$ and $\vM$, we need to estimate them using the prior structural information of $\mathbf{M}$. To this end, the proposed QPMA algorithm is comprised of three stages: (i) estimating the column space of $\mathbf{M}$; (ii) followed by estimating the unknown polynomial coefficient matrix, $\mathbf{Q}$, and subsequently estimating the row space of $\mathbf{M}$ by leveraging the quasi polynomial structure; and (iii) the final matrix approximation step constrained to the row and column space approximations obtained previously. The complete algorithm is summarized in Algorithm~\ref{alg:alg}.

We first estimate the column space of $\mathbf{M}$ using  $\mathbf{A} = \mathbf{M} \mathbf{\Psi}$. We argue that as long as enough independent columns are sampled (this is shown in Lemma~\ref{eq:lem1}), the following optimization gives us a good estimate
\begin{align*}
    \mathbf{U}_A = \argmin_{\tilde{\mathbf{U}} \in \RR^{n \times r}, \tilde{\mathbf{U}}^\transp\tilde{\mathbf{U}} = \mathbf{I}} \left\Vert(\mathbf{I} - \tilde{\mathbf{U}}\tilde{\mathbf{U}}^\transp)\mathbf{A} \mathbf{A}^\transp\right\Vert_2.
\end{align*}
From Eckart-Young-Mirsky theorem, the solution to the above is given by the rank-$r$ SVD of $\mathbf{A}$ (the matrix formed by the left singular vectors corresponding to the top-$r$ singular values),
\begin{align}\label{eq:svd_A}
    \mathbf{A}\stackrel{\text{r-SVD}}{=}\mathbf{U}_{ A}\mathbf{\Sigma}_{A}\mathbf{V}_{ A}^{\trasp}.
\end{align}

\begin{algorithm}[ht!]
\caption{{Quasi-Polynomial Matrix Approximation}}\label{alg:alg}
\begin{algorithmic}[1]
\State {\bf Input:} $\mathbf{A} \in \RR^{n \times d}$(A matrix of sampled columns), $\mathbf{S} \in \RR^{l \times m}$(A polynomial basis matrix), {{$\mathbf{\Psi}$ (Column sampling operator)}}
\State {\bf Parameters:} A target rank $r$, Degree of polynomial $l$, Step size $\eta$, Max iteration $T$
\State {\bf Initialization:} 
{{generate each entry of $\hat{\mathbf{Q}}_{1}$ independently from $\mathcal{N}(0,1)$}}
\State { \; {\bf Column space estimation}}\\ \;\;\; Do rank-$r$ SVD of $\mathbf{A}$ as $\mathbf{A}\overset{r-\mathrm{SVD}}{=}\mathbf{U}_{A}\mathbf{\Lambda}_{A}\mathbf{V}_{A}^{\trasp}$

\State {\bf \;\; Row space estimation} \\ \;\;\;\; For $t\in[T]$, do
    \item[]\;\;\;\;\;\;\;\;$\hat{\mathbf{Q}}_{t+1} = \hat{\mathbf{Q}}_{t} - \eta \left(\mathbf{A}-\hat{\mathbf{Q}}_{t}\mathbf{S}\mathbf{\Psi}\right)\left(\mathbf{S}\mathbf{\Psi}\right)^{\trasp}$ \\ \;\;\;\; With $\hat{\mathbf{ Q}}\equiv\hat{\mathbf{Q}}_{T}$, compute $\hat{\mathbf {Q}}{\mathbf{ S}}\overset{r-\mathrm{SVD}}{=}\hat{\mathbf{U}}_{QS}\hat{\mathbf{\Lambda}}_{QS}\hat{\mathbf{V}}^{\trasp}_{QS}$

\State {\bf \;\; Matrix approximation}\\
\;\;\;\; Using ${\mathbf{U}}_{A}$ and $\vqsh$, for $t\in[T]$, do
    \item[]\;\;\;\;\; $\hat{\mathbf{ Z}}_{t+1}
    = \hat{\mathbf{ Z}}_{t} - \eta {\mathbf{U}}_{A}^{\trasp}\left(\mathbf{A}-{\mathbf{U}}_{A}\hat{\mathbf{Z}}_{t}\vqsh^{\trasp}{\bf\Psi}\right)\left(\vqsh^{\trasp}{\bf\Psi}\right)^{\trasp}$\\ \;\;\;\; Obtain $\hat{\mathbf{ Z}}=\hat{\bf Z}_{T}$ \\ \;\;\;\; Complete $\hat{\mathbf{M}}={\bf U}_{A}\hat{\mathbf{ Z}}\vqsh^{\trasp}$
\State {\bf Output:} $\hat{\mathbf{M}}=\mathbf{U}_{A}\hat{\mathbf{Z}}\vqsh^{\trasp}$.
\end{algorithmic}
\end{algorithm}

We next estimate the unknown polynomial coefficient matrix, $\mathbf{Q}$ as follows
\begin{align}\label{eq:opt_Q}
\hat{\mathbf{Q}}=\argmin_{\tilde{\mathbf{Q}}}\;\; {\left\Vert  \mathbf{A}- \tilde{\mathbf{Q}}{\mathbf{S}}{\mathbf{\Psi}} \right\Vert}_{F}^{2}.
\end{align}
This is a standard regression problem that admits a closed form solution, but it is computationally expensive to compute a pseudo-inverses. Thus, we instead consider a gradient descent approach \cite{boyd2004convex}. Concretely, we define $g(\mathbf{Q}) = {\left\Vert  \mathbf{A}- {\mathbf{QS}\mathbf{\Psi}} \right\Vert}_{F}^{2}$. The gradient of $g(\mathbf{Q})$ with respect to $\mathbf{Q}$ is given by 
\begin{align}
    \triangledown_\mathbf{Q} g(\mathbf{Q}) = 2\left(\mathbf{A}-{\mathbf{QS}\mathbf{\Psi}}\right)\left(\mathbf{S}\mathbf{\Psi}\right)^{\trasp}.\nonumber
\end{align} 
We then repeat the following update rule at each iteration $t = [T]$ until convergence: 
\begin{align}\label{eq:update_q}
    \hat{\mathbf{Q}}_{t+1} = \hat{\mathbf{Q}}_{t} - \eta \left(\mathbf{A}-\mathbf{\hat{Q}}_{t}\mathbf {S}\mathbf{\Psi}\right)\left(\mathbf{S}\mathbf{\Psi}\right)^{\trasp},
\end{align} where $\eta$ is an appropriately chosen step size.
Now if $\hat{\mathbf{Q}} \equiv \hat{\mathbf{Q}}_{T} $ is a good approximation of $\mathbf{Q}$ (this is shown in Lemma~\ref{eq:lem3}), we can obtain the row space information of $\mathbf{M}$ through the following minimization
\begin{align}\label{eq:estimated_qs}
    \vqsh = \argmin_{\tilde{\mathbf{V}} \in \RR^{m \times r}, \tilde{\mathbf{V}}^\transp\tilde{\mathbf{V}} = \mathbf{I}} \left\Vert(\mathbf{I} - \tilde{\mathbf{V}}\tilde{\mathbf{V}}^\transp)(\mathbf{\hat{Q}S})^\transp (\mathbf{\hat{Q}S})\right\Vert_2
\end{align}
Again, the solution to the above is readily obtained through a rank-$r$ SVD of $\hat{\mathbf{Q}}\mathbf{S}$. 

Finally, we exploit the row and column space estimates to obtain the low-rank approximation as follows
\begin{align}\label{eq:opt_Z}
 \hat{\mathbf{Z}} & = \arg   \min_{\tilde{\mathbf{ Z}}}\;\;  \left\Vert \mathbf{A}-{\mathbf{U}}_{A}\tilde{\mathbf{Z}}\vqsh^{\trasp}{\bf\Psi}\right\Vert_{F}^{2}.
\end{align} 
We observe that \eqref{eq:opt_Z} is also a regression problem that we solve through an gradient descent method. Define 
\begin{align}\label{eq:opt_Z_final}
    f(\mathbf{Z})=\left\Vert \mathbf{A}-{\mathbf{U}}_{A}\mathbf{Z}\vqsh^{\trasp}{\mathbf{\Psi}}\right\Vert_{F}^{2}.    
\end{align} 
The gradient of $f({\mathbf{Z}})$ is given by
\begin{align}
    \triangledown_{\mathbf{ Z}} f(\mathbf{Z}) = 2{\mathbf{U}}_{A}^{\trasp}\left(\mathbf{A}-{\mathbf{U}}_{A}\mathbf{Z}\vqsh^{\trasp}{\mathbf{\Psi}}\right)\left(\vqsh^{\trasp}{\mathbf{\Psi}}\right)^{\trasp}\nonumber,
\end{align}
This finally yields the reconstructed matrix 
\begin{align}\label{eq:hat_M}
    \hat{\mathbf{M}} = {\mathbf{U}}_{A}\hat{\mathbf{Z}}\vqsh^{\trasp},
\end{align} 
This concludes the algorithm.


}

\section{Main Result and Proof Sketch}
In this section, we provide our main result and the proof sketch. We require the following definitions before presenting the main result.
We consider the following standard definition of matrix incoherence \cite{candes_exact_2009}. 

\begin{definition}[Incoherence] 
\label{def:incoh} Let $\mathbf{X}$ be a ${n \times m}$ matrix of rank $r$ and $\mathbf{X}\stackrel{\text{$r$-SVD}}{=}\mathbf{U}\mathbf{\Sigma}\mathbf{V}^{\trasp}$. Let $\mathbf{u}_{i}$ be the $i$-th row of $\mathbf{U}$ and $\mathbf{v}_{j}$ be the $j$-th row of $\mathbf{V}$. Then, the incoherence of $\mathbf{X}$ is given by
\begin{align}
    &\mu(\mathbf{X})\nonumber
    = \max\left(\max_{i\in[n]}\frac{n}{r}{\Vert \mathbf{u}_{i} \Vert}^2_2,\max_{j\in[m]}\frac{m}{r}{\Vert \mathbf{v}_{j} \Vert}^2_2\right)\nonumber.
\end{align}
\end{definition}

Incoherence is a necessary assumption to ensure the ``energy'' is spread out uniformly to complete a matrix from a few randomly chosen entries. 
We note that despite the fact that our work deals with the setting wherein a few randomly chosen {\em columns} are observed (as opposed to a few randomly chosen {\em entries} that standard MC studies), the inclusion of the quasi-polynomial side information allows us to work with the standard incoherence definition. In our analysis, we use the shorthand notation, $\mu\doteq\mu(\mathbf{M})$ and $\hat{\mu}\doteq\mu(\hat{\mathbf{M}})$ \footnote{In our current result, we assume that the output of Algorithm \ref{alg:alg} is incoherent. We will consider eliminating this assumption as part of future work.}.

Next, we review strong convexity of a function \cite[sec 3]{boyd2004convex}.  
\begin{definition}[Strong Convexity]\label{def:strong_convexity} A differentiable\footnote{If $f$ is non-differentiable, then the gradient of $f$ is replaced by its sub-gradient.} function $f: \RR^{n} \rightarrow \RR$ is strongly convex with parameter $\alpha>0$ if the following holds for all $\mathbf{x}, \mathbf{x'} \in \RR^{n}$, 
\begin{align}
    f(\mathbf{x}) \geq f({\mathbf{x'}}) + \triangledown f(\mathbf {x'})({\mathbf{x}} - {\mathbf{x'}}) + \frac{\alpha}{2}{\Vert {\mathbf{x}}-{\mathbf{x'}} \Vert}^2_2.\nonumber
\end{align}
\end{definition}

We use Definition \ref{def:strong_convexity} to derive convergence guarantees for the row space estimation, and the matrix approximation steps of QPMA.

\subsection{Main Result}

We need the following assumption before presenting our main result.

\begin{assumption}\label{as:as1} Let $\delta_1 := \sigma_r(\mathbf{M-E}) - \sigma_{r+1}(\mathbf{M}) > 0$.
\end{assumption}



Observe that $\delta_1$ captures the effective eigengap of $\mathbf{M}$. We now present our main result.


\renewcommand{\SNR}{\mbox{{SNR}}}

{{
\begin{theorem}\label{eq:thm1} Consider measurements that satisfy Assumption~\ref{as:as1}. 
Assume that $d$ columns are sampled uniformly at random from the underlying ground truth, $\mathbf{M}$. Then, if $d \geq  \mbox{max} \left\{c_1 {\mu}r\ln r, c_{2}{\hat{\mu}}^{2}r^{2}\ln r \right\}$, with probability at least $1-c_3 r^{-10}$ we have
\begin{align}\label{eq:main_res_thm1}
\frac{{\Vert\mathbf{M}- \hat{\mathbf{M}}\Vert}_{2}^{2}}{\|\mathbf{M}\|_2^2} &\leq
4 \frac{\sigma_{r+1}^2(\mathbf{M})}{\sigma_1^2(\mathbf{M})} \left( 1 + \frac{4m}{d} \right) \left( 3 + \frac{n}{d} \right) \nonumber \\
& + 64 \|\mathbf{E}\|_F^2 \left( 1 + \frac{4m}{d}\right) \left( \frac{1}{\delta_1^2} + \frac{1}{\delta_s^2} \right)
\end{align}
where $\delta_s : = \frac{\sigma_l(\mathbf{QS})}{\|(\mathbf{S\Psi})^{\dagger}\mathbf{S}\|_F}$ and $c_1, c_2, c_3 >0$ are numerical constants. 
\end{theorem}}}

\begin{proof}[Proof]
Theorem \ref{eq:thm1} is proved in the Appendix. The proof follows from applying large-deviation style results from random matrix theory \cite{tropp2011improved} to ensure that the loss-function in \eqref{eq:opt_Z_final} is well-behaved as long as we sample a sufficient number of columns, followed by a careful application of Wedin's theorem \cite{stewart1998perturbation}.  
\end{proof}

If $\mathbf{E} = \mathbf{0}$, we have the following Corollary.

\begin{corollary}[Perfect Side-Information] \label{cor:noiseless} 
Under the conditions of Theorem \ref{eq:thm1}, if $\mathbf{E} = \mathbf{0}$, then with probability at least $1 - Cr^{-10}$, where $C>0$, 
{{
\begin{align}
{\left\Vert{\mathbf{ M}}-\hat{\mathbf{M}}\right\Vert}_{2}^{2} & \lesssim\frac{80mn}{d^2}\sigma^{2}_{r+1}(\mathbf{M}) = O\left({\frac{mn}{d^2}}{\left\Vert\mathbf{M}-{ {\mathbf{M}}_{r}}\right\Vert}_{2}^2\right)
\end{align} }}
where ${ {\mathbf{M}}_{r}}$ is the best rank-$r$ approximation of $\mathbf{M}$.
\end{corollary}

\newcommand{\rank}{\mathrm{rank}}
\newcommand{\M}{\mathbf{M}}
\subsection{Discussion}\label{sec:disc} 
{\subsubsection{Interpreting the Signal-to-Noise Ratio}
Recall from Assumption \ref{as:as1} that $\delta_1$ captures the effective eigengap of $\mathbf{M}$ and thus a natural interpretation of the term $\frac{\delta_1^2}{\|\mathbf{E}\|_F^2}$ is the ``signal-to-noise ratio''$(\mbox{SNR})$. Furthermore, we observe from Theorem \ref{eq:thm1} that $\delta_s : = \frac{\sigma_l(\mathbf{QS})}{\|(\mathbf{S\Psi})^{\dagger}\mathbf{S}\|_F}$ essentially measures {\em how informative} the side information, $\mathbf{S}$ is. To be more precise, first consider the numerator term $\sigma_l(\mathbf{QS})$. The larger this quantity, the more dominant (w.r.t $\mathbf{E}$) and hence, the more informative, the side-information is. The denominator term, $\|(\mathbf{S\Psi})^\dagger\mathbf{S}\|$, on the other hand, measures how much of the side information is effectively captured after the column-sampling process. Observe that if $\mathbf{\Psi} = \mathbf{I}$, then $\|(\mathbf{S\Psi})^{\dagger}\mathbf{S}\|_F = \|(\mathbf{S})^{\dagger}\mathbf{S}\|_F = \sqrt{l}$, and as expected, this value reduces as the number of sampled columns, $d$, reduces. Finally, we emphasize that from the perspective of the motivating application, we can control $\mathbf{S}$ and thus, it is possible to ensure that $\|(\mathbf{S\Psi})^{\dagger}\mathbf{S}\|_F = O(1)$. With a slight abuse of terminology, we use
\begin{align*}
    \mbox{SNR} := \frac{1}{\|\mathbf{E}\|_F} \left( \frac{1}{\delta_1^2} + \frac{1}{\delta_s^2} \right)^{-1/2}
\end{align*}
in the sequel.
}

In Theorem \ref{eq:thm1}, we focus on the two sources of error: (i) the {\em unrecoverable energy} that arises due to fact that the original matrix is high-rank; and (ii) the {\em imperfect side-information}. The first term in \eqref{eq:main_res_thm1} represents the unrecoverable energy, as we seek a low-rank approximation of a high-rank matrix. Even if we had perfect side information, i.e., $\mathbf{E} = \mathbf{0}$ there will be an error incurred due to the low-rank approximation. We also observe that QPMA suffers a multiplicative factor of $O(m/d)$ coupled with the best rank-$r$ approximation error, {$\|\M - \M_r\|_{2} = \sigma_{r+1}(\M)$.} This is owing to the fact that we solve a harder problem than classical rank-$r$ approximation and this multiplicative factor is standard in the high-rank matrix approximation literature \cite{matrix_recovery_sketch,eckart1936approximation,mahoney2009cur, cs_covariance}. {{The second term in \eqref{eq:main_res_thm1} occurs due to the imperfect nature of the side information, i.e., since $\mathbf{E} \neq \mathbf{0}$.}} 
We emphasize that since our main result does not assume any statistical or generative models on the noise, it is highly non-trivial to make further deductions.  Thus, we consider specific noise models, and the side-information matrices in future work.  

\subsubsection{Comparison with CUR+ \cite{xu2015cur}}
We assume for the rest of the paper that the incoherences, are constant\footnote{In this paper, we use the order notation with respect to $m, n$.}, i.e., $\mu, \hat{\mu} = O(1)$. With this, it is easy to see from Theorem \ref{eq:thm1} that {{ $d=O(r^2 \ln r)$. Observe that in order to obtain a non-trivial rank $r$ approximation, one needs to sample at least $r$ columns of $\M$ even with perfect side information. 
Theorem \ref{eq:thm1} shows that with mismatched side-information and unstructured noise, QPMA obtains a good approximation with just $O(r^2 \ln r)$ columns}}.  We contrast with CUR+ since its sampling structure is the most similar to our problem setting and is also the state-of-the-art in high-rank matrix approximation with incomplete measurements. As opposed to QPMA, CUR+ requires a $d= O(r \ln r)$ randomly chosen rows {and} columns, and an additional $O(r^2 \ln r)$ randomly chosen entries. Thus, by imposing a significantly weaker assumption: a quasi-polynomial side information instead of observing a subset of rows, QPMA attains a sample complexity bound that is a factor of $r$ worse than that of CUR+. We believe that this bound can be improved by a more refined proof technique for Lemma \ref{eq:lem3} which we defer to future work.


\subsubsection{Interpreting the Error}
Many prior error analyses for high-rank matrix approximation \cite{mahoney2009cur,drineas2008relative} have the following common structure for the error bound:
\begin{align}
    \left\Vert \mathbf{M}-\hat{\mathbf{M}}\right\Vert \leq (1 + \epsilon_1)\left\Vert \mathbf{M}-\mathbf{M}_{r}\right\Vert +\epsilon_2\left\Vert \mathbf{M}\right\Vert,
\end{align} 
where $\mathbf{M}_{r}$ is the best rank-$r$ approximation of a matrix $\mathbf{M}$ and $ \epsilon_1 >0$, and $\epsilon_2 \in (0,1)$ are derived constants that are specialized to the problem. We see that we can formulate our error bound from Theorem \ref{eq:thm1} in a similar fashion,
\begin{align}
    \left\Vert \mathbf{M}-\hat{\mathbf{M}}\right\Vert &\leq  O\left(\frac{m}{d}\left\Vert \mathbf{M}-\mathbf{M}_{r}\right\Vert\right)  + O\left(\sqrt{\frac{m}{d}} \mbox{SNR}^{-1} \|\M\|\right) \nonumber
\end{align}
where, without loss of generality, we assume $m \geq n$. As previously mentioned, the scaling factor $\frac{m}{d} (\equiv 1 + \epsilon_1)$ is, in general, unavoidable due the high-rank nature of the true matrix in addition to sub-sampling of the columns. We also notice that $\epsilon_2 = C \sqrt{\frac{m}{d}}\cdot\frac{1}{\mbox{SNR}}$, i.e., the noise is amplified by the square root of the sub-sampling factor, $\sqrt{\frac{m}{d}}$ as well. We emphasize that unlike results in PCA, wherein there is a ``denoising'' effect with increasing the number of observations, matrix approximation algorithms do not possess the ability to denoise, the observations. However, as expected, increasing the number of observed columns reduces the approximation error and approaches the best-case scenario of $C\mbox{SNR}^{-1}$ as $m \to d$. Finally, we mention that in the setting where $\|\mathbf{E}\|_{F} = 0$,  and if the matrix is ``roughly square'', i.e., $ m = O(n)$, our result improves upon CUR+ \cite[Theorem 2]{xu2015cur} by a factor of $\sqrt{m}$. 


\begin{figure*}[t!]
\centering
    \includegraphics[width=18cm]{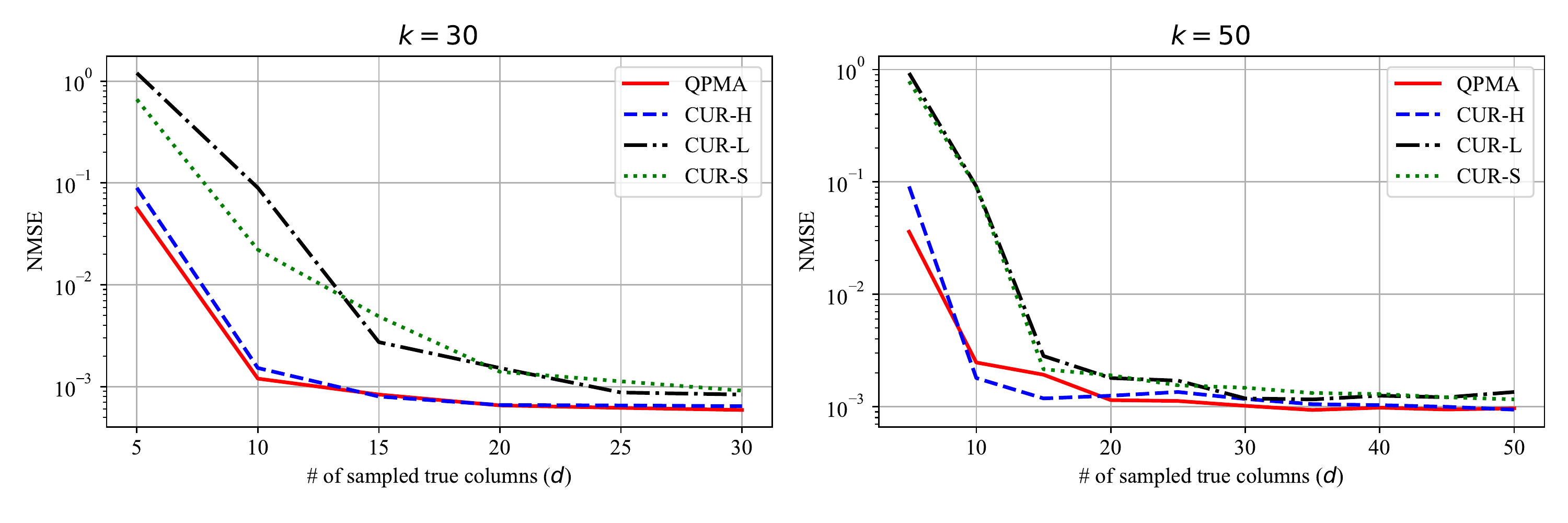}
   \vspace*{-0.2in}
    \caption{The NMSE versus the number of true sampled columns, $d$. Here, $k=\{30, 50\}$ and $l=5$.}
    \label{fig:perform6}
\end{figure*}

{{\subsubsection{Time complexity of QPMA} {We next derive the computational complexity of Algorithm \ref{alg:alg}. The column space $\mathbf{U}_A$ is estimated through a rank-$r$ SVD on $\mathbf{A}$, and this takes $O(ndr)$ time \cite{brand2006fast}\footnote{Note that we only require the top-$r$ singular vectors, but do not require the singular values, and hence there is no dependence on the singular value gap}. Next, the row space estimation step is performed by first estimating the polynomial coefficient matrix $\mathbf{Q}$ by gradient descent (GD). The run time for the corresponding matrix multiply in each iteration is $O(\max(nl^2d, nd^2l)) = O(nd^2l)$ (we assume $d >l$ without loss of generality) and thus the overall complexity of GD is 
$O(nd^2l)$ given a bounded gradient assumption and a bounded initial error\footnote{In this paper, we assume that the number of iterations for the GD step is $O(1)$. We do this since without additional statistical assumptions on the signal model, characterizing $T$ is very complex, and beyond the scope of this paper} \cite[Sec 9]{boyd2004convex}. 
Next, the rank-$r$ SVD  of $\hat{\mathbf{Q}}\mathbf{S}$ can be performed in $O(n m r)$ time. Finally, the per-iteration complexity for the matrix approximation step is $O(\max(n^2dr, nd^2r)) = O(n^{2}dr)$ and since we assume that the number of iterations, $T = O(1)$, the overall running time for GD is $O(n^2dr)$. 
Thus, the overall computational complexity of Algorithm \ref{alg:alg} is $O(\max(nmr, n^2dl))$ that is equal (up to constant factors) to performing a rank-$r$ SVD on the original matrix, $\mathbf{M}$.}}}

\subsection{Proof Sketch and key Lemmas}
Here we provide the proof sketch and the main Lemmas required to prove Theorem \ref{eq:thm1}. 
The complete proofs are provided in Appendix.

We first bound the error as ${\Vert {\bf M}-\hat{\bf M} \Vert}^2_2$ as 
\begin{align}\label{eq:eq7}
    \left\Vert\mathbf{M}- \hat{\mathbf{M}}\right\Vert_{2}^{2}&
    = \left\Vert \mathbf{M}-{\mathbf{ U}_{A}}\hat{\mathbf{Z}}\vqsh^{\trasp} \right\Vert_{2}^{2}\nonumber\\
    &= \Vert\mathbf{M}-\mathbf{ P}_{\mathbf{ U}_{A}}\mathbf{M}\mathbf{ P}_{\vqsh} \nonumber\\
    &\;\;\;\;\;\;\;\;\;+\mathbf{P}_{\mathbf{U}_{A}}\mathbf{M}\mathbf{P}_{\vqsh}-\mathbf{U}_{A}\hat{\mathbf{ Z}}\vqsh^{\trasp}\Vert_{2}^{2} \nonumber\\
    &\stackrel{\text{(a)}}{\leq}2\underbrace{\left\Vert\mathbf{M}-\mathbf{ P}_{\mathbf{ U}_{A}}\mathbf{M}\mathbf {P}_{\vqsh}\right\Vert_{2}^{2}}_{\odot}\nonumber\\
    &\;\;\;\;+ 2\underbrace{\left\Vert\mathbf{ P}_{{\mathbf{ U}_{A}}}{\mathbf{M}}\mathbf{P}_{\vqsh}- \mathbf{ U}_{A}\hat{\mathbf{Z}}\vqsh^{\trasp}\right\Vert_{2}^{2}}_{\circledcirc}
\end{align} 
where (a) follows from the triangle inequality and the fact that for $a,b \geq 0$, $(a+b)^2 \leq 2(a^2 + b^2)$. Next, recall that $\mathbf{P}_{{\mathbf{U}_{A}}}={\mathbf{U}_{A}{\mathbf{U}}_{A}^{\trasp}}$ and $\mathbf{ P}_{\vqsh}=\vqsh\vqsh^{\trasp}$. Notice that can obtain high probability bounds on $\odot$ and $\circledcirc$, we are done. To that end, we first consider $\odot$. 

Note that $\odot$ captures the energy of the true matrix, $\M$ orthogonal to the estimated ($r$-dimensional) row and column spaces. We provide a bound for this below in Lemma \ref{eq:lem1}.

{{
\vspace{0.01cm}
\begin{lemma}\label{eq:lem1}
Consider measurements that satisfy Assumption~\ref{as:as1}. Then, if $d \geq c_1\mu r \ln r$, under the conditions of Theorem \ref{eq:thm1}, with probability at least $1 - c_2r^{-10}$ we have that 
\begin{align}\label{eq:theo2}
   &{\left\Vert \mathbf{ M}-\mathbf{P}_{\mathbf{U}_{A}}\mathbf{M}\mathbf{P}_{\vqsh}\right\Vert}_{2}^{2}\nonumber \\
   &\;\;\;\leq 2\sigma^{2}_{r+1}\left(\mathbf{M}\right)\left(3+\frac{n}{d}\right)+32\sigma_{1}^{2}(\mathbf{M})\left\Vert\mathbf{E}\right\Vert_{F}^{2}\left(\frac{1}{\delta_1^2}+ \frac{1}{\delta_s^2}\right)
\end{align}
where $\delta_s := \frac{\left\Vert\left(\mathbf{S\Psi}\right)^{\dagger}\mathbf{S}\right\Vert_{F}}{\sigma_{l}(\mathbf{QS})}$ and $c_1, c_2 >0$ are numerical constants.
\end{lemma}}}

\begin{figure*}[t!]
    \centering
    \includegraphics[width=18cm]{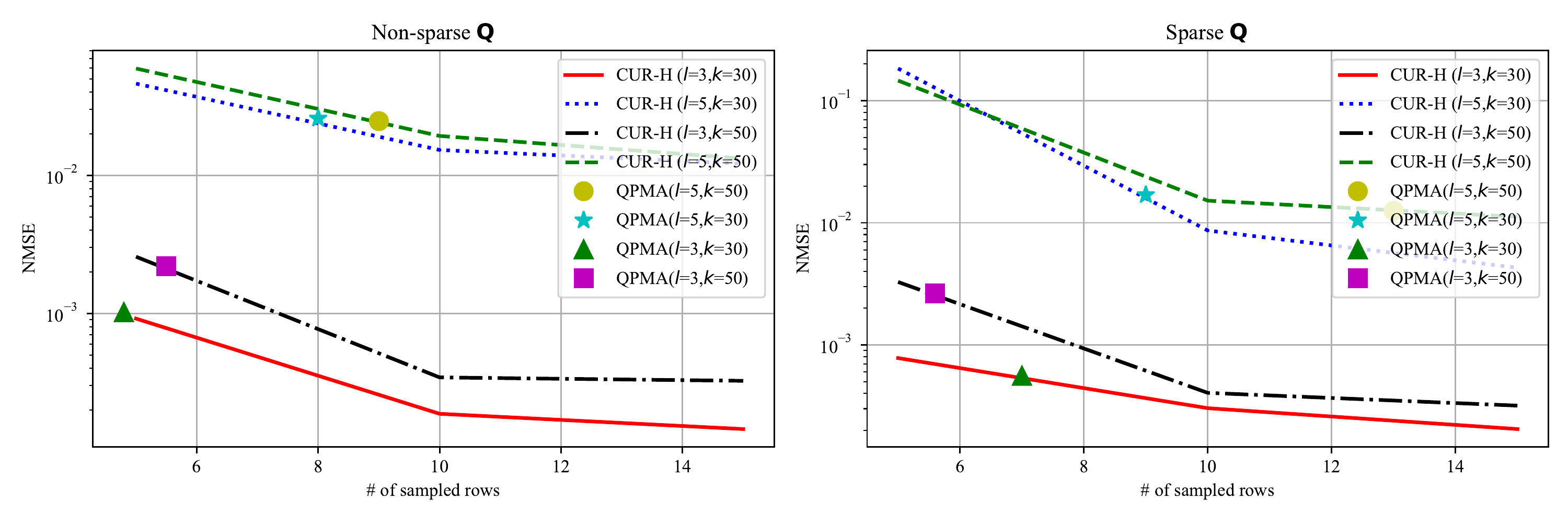}
        \vspace*{-0.1in}
    \caption{Characterizing the informativeness of the quasi-polynomial side information through numerical performance of CUR+ when $d=5$ and $r=5$}
    \label{fig:p4}
\end{figure*}

The complete proof of Lemma \ref{eq:lem1} is provided in Appendix \ref{sec:appendix_a}. 
The proof follows from first invoking \cite[Theorem 6]{xu2015cur_sup} to bound the energy of $\M$ orthogonal to $\mathbf{U}_A$ and $\hat{\mathbf{V}}_{QS}$, where $\hat{\mathbf{V}}_{QS}$ is the row space of $\mathbf{\hat{Q}S}$, followed by a careful application of Wedin's Theorem \cite{stewart1998perturbation} (provided in appendix as Theorem \ref{def:wedin}) to bound the ``distance'' between {{ the ``true row space'' ${\mathbf{V}}_{QS}$ defined in \eqref{eq:svd_qs} and the estimated  $\hat{\mathbf{V}}_{QS}$ obtained from \eqref{eq:estimated_qs}.}} These are provided as Lemmas \ref{eq:lem4} and \ref{eq:lem5} respectively. 

Akin to the result of \cite[Theorem 2]{xu2015cur} and as explained previously, Lemma \ref{eq:lem1}, consists of error due to the fact that $\rank(\M) = k \gg r$ and sub-sampling of columns (both contribute to the first term). When $k=r$, the first term is zero since $\sigma_{r+1}(\mathbf{M}) = 0$.  The second term corresponds to the error due to noise and imperfect nature of side information.



Next, $\circledcirc$ essentially captures the error in the final matrix approximation step, estimation of $\mathbf{Z}$. This is bounded using Lemma \ref{eq:lem3} below. 


{{
\begin{lemma}\label{eq:lem3} 
Consider measurements that satisfy Assumption~\ref{as:as1}. Let the objective function in \eqref{eq:final_opt} be $\alpha$-strongly convex with $\alpha \geq d/2m$. Then, if $d \geq c_{2}{\hat{\mu}}^{2}r^{2}\ln r$, under the conditions of Theorem \ref{eq:thm1}, with probability at least $1 - r^{-c_1}$ we have that
\begin{align}
&\Vert\mathbf{P}_{\mathbf{U}_{A}}\mathbf{ M}\mathbf{P}_{\vqsh}-\mathbf{U}_{A}\hat{\mathbf{ Z}}\vqsh^{\trasp}\Vert_{2}^{2}\nonumber\\ &\leq
\frac{4m}{d}\left[2\sigma^{2}_{r+1}\left(\mathbf{M}\right)\left(3+\frac{n}{d}\right)+32\sigma_{1}^{2}(\mathbf{M})\left\Vert\mathbf{E}\right\Vert_{F}^{2}\left(\frac{1}{\delta_1^2}+ \frac{1}{\delta_s^2}\right)\right]\nonumber,
\end{align}
where $\delta_s := \frac{\left\Vert\left(\mathbf{S\Psi}\right)^{\dagger}\mathbf{S}\right\Vert_{F}}{\sigma_{l}(\mathbf{QS})}$ and $c_1, c_2 >0$ are numerical constants.
\end{lemma}}}

The proof of Lemma \ref{eq:lem3} is provided in Appendix \ref{sec:appendix_c} and the proof follows by leveraging the fact that the objective function, $f(\mathbf{Z})$, in \eqref{eq:opt_Z_final} is $\alpha$-strongly convex with $\alpha \geq d/2m$, the result of Lemma \ref{eq:lem2} and some simple linear algebra tricks.

We next show that the objective function in \eqref{eq:opt_Z_final}, $f(\mathbf{Z})$  is indeed strongly convex with the requisite parameter setting in Lemma \ref{eq:lem2}.

{{
\begin{lemma}\label{eq:lem2}  
Under the conditions of Theorem \ref{eq:thm1}, with probability at least $1 - r^{-c_1}$,  the objective function, $f(\mathbf{Z})$ in \eqref{eq:opt_Z_final}, is $\alpha$-strongly convex with $\alpha \geq \frac{d}{2m}$ as long as 
\begin{align}
    d \geq c_{2}{\hat{\mu}}^{2}r^{2}\ln r.\nonumber
\end{align} 
where $c_1, c_2 >0 $ are numerical constants.
\end{lemma}}}

Intuitively, Lemma \ref{eq:lem2} shows that as long as the number of columns is large enough, the objective function for gradient descent has a quadratic lower bound on the curvature. The proof follows a careful application of a large-deviation result for sums of random matrices \cite{tropp2011improved} followed by linear algebraic computation.

Combining Lemmas \ref{eq:lem1}, \ref{eq:lem3} and \ref{eq:lem2} completes the proof.

\vspace{0.2cm}

\section{Numerical Results}
\label{sec:exp}
Herein, we investigate QPMA's performance on both synthetic and real-world data. All experiments on synthetic data are averaged over 100 independent iterations. The code can be found at \url{https://github.com/JeongminChae/QPMA.} 

\subsubsection{Benchmark Algorithms}
A challenge in finding comparison strategies is that, as previously noted, matrix approximation algorithms typically require row and column space information, provided by the samples.  We compare QPMA (Algorithm \ref{alg:alg}) with three variants of CUR+ \cite{xu2015cur} which are generated based on different sampling strategies as outlined in Table \ref{tb:tb3}. As noted in Section~\ref{sec:intro}, CUR+ samples a subset of the full columns and  full rows, as well as additional random entries. Comparing with QPMA in Algorithm~\ref{alg:alg}, CUR+ does row space estimation via rank-$r$ SVD of the sampled true rows. And thus, for CUR+, the matrix approximation step correspoinding to line 9 is performed with the column space and row space estimated with the true columns and rows. In contrast, QPMA estimates the row space, exploiting the side information, but no sampled rows.

QPMA employs column sampling only from $d$ columns and thus utilizes $nd$ samples in total.  CUR-S employs the same number of samples as QPMA with column, row and random sampling.  CUR-L employs a reduced number of total samples with $\frac{d}{2}$ rows and columns each, but no randomized sampling.  Similarly, CUR-H employs an increased number of samples relative to QPMA with $d$ rows and columns each and also no additional randomized sampling.

\begin{table}[ht!]
    \caption{Comparison of Sampling Schemes for Baseline Algorithms}
    \begin{center}
    \begin{tabular}{c|c|c|c|c}
    \toprule
    Algorithm & \# rows & \# columns  & random entries & Total samples \\  \hline
    CUR-L & $d/2$ & $d/2$ & 0 & $nd-d^{2}/4$\\ \hline
    CUR-S & $d/2$ & $d/2$ & $d^{2}/4$ & $nd$\\ \hline
    CUR-H & $d$ & $d$ & 0 & $2nd-d^{2}$\\ \hline
    {\bf QPMA} & 0 & $d$ & 0 & $nd$\\
    \bottomrule
    \end{tabular}
    \label{tb:tb3}
    \end{center}
\end{table}




\subsection{Synthetic Data} 


\subsubsection{Varying $d$} 
We first investigate  performance as a function of the number of samples, governed by $d$.  We generate the data as follows: The entries of the polynomial coefficient matrix, $\mathbf{Q} \in \RR^{n \times l}$ are drawn i.i.d. from $\mathcal{N}(0,1)$. We generate the reaction coordinate values, $\mathbf{s} \in \RR^m$ as $[1 + 0.01 *[m]]^\transp = [1.01, 1.02, \ \cdots, 1 + 0.01m]^\transp$ and subsequently, the side information matrix $\mathbf{S}$ as in \eqref{eq:S}. For all experiments, we set $n = 100$ and $m = 100$. Next, in order to simulataneously control the ``noise level'' and the rank of the true matrix, $\mathbf{M}$, we generate the perturbation matrix, $\mathbf{E}$ as follows. Recall that $\mathbf{QS} \overset{\mathrm{SVD}}{=} \mathbf{\tilde U} \mathbf{\tilde \Sigma} \mathbf{\tilde V}^\transp$. We set $\mathbf{E} = \mathbf{\tilde U}_{[k]} \mathbf{\tilde R} \mathbf{\tilde V}_{[k]}^\transp$, where $\mathbf{\tilde U}_{[k]} \in \RR^{n\times k}$ the matrix of the first $k$ columns of $\mathbf{\tilde U}$ and similarly for $\mathbf{\tilde V}_{[k]}$. The entries of $\mathbf{\tilde R} \in \RR^{k \times k}$ are drawn i.i.d. from $\mathcal{N}(0, \sigma^2)$. In the first experiment, we consider two values of the true rank $k = \{30, 50\}$, two possible polynomial degrees $l = \{3,5\}$, and a noise standard deviation of $\sigma = 10^{-4}$. We observe that each entry in $\mathbf{E}$ has a standard deviation of $\sigma$ and thus, the aggregated noise is much higher, as our theoretical guarantees (and numerical results) are shown relative to the Frobenius norm error. 

We implement Algorithm \ref{alg:alg} with {fixed step-size $\eta=0.01$, and the maximum number of iterations $T = 1500$.} We implement all variants of CUR+ with default parameters and we set $T=1500$ to provide a fair comparison with Algorithm \ref{alg:alg}. We plot the normalized mean square error (NMSE), $\|\mathbf{M - \hat{M}}\|_F/\|\mathbf{M}\|_F$ for all algorithms in Fig. \ref{fig:perform6}. We notice that for both values of $k$, despite observing much fewer samples than CUR-H, the performance of QPMA is comparable to that of CUR-H. Furthermore, both CUR-H and QPMA uniformly outperforms the lower sample variants of CUR+ (CUR-L and CUR-S). This suggests that Algorithm \ref{alg:alg} effectively exploits the quasi-polynomial side information. Furthermore, in the low-sample regime, i.e., $d \leq 10$, {both QPMA and CUR-H are almost two/three orders of magnitude better than CUR-S and CUR-L. }Finally, we notice that as the number of columns, $d$, approaches the true rank, $k$, the performance of all algorithms do not significantly improve. This is in agreement with Theorem \ref{eq:thm1} since in this regime, the error is dominated by the presence of noise, $\mathbf{E}$.

\begin{figure}[t!]
    \centering
    \includegraphics[width=9cm]{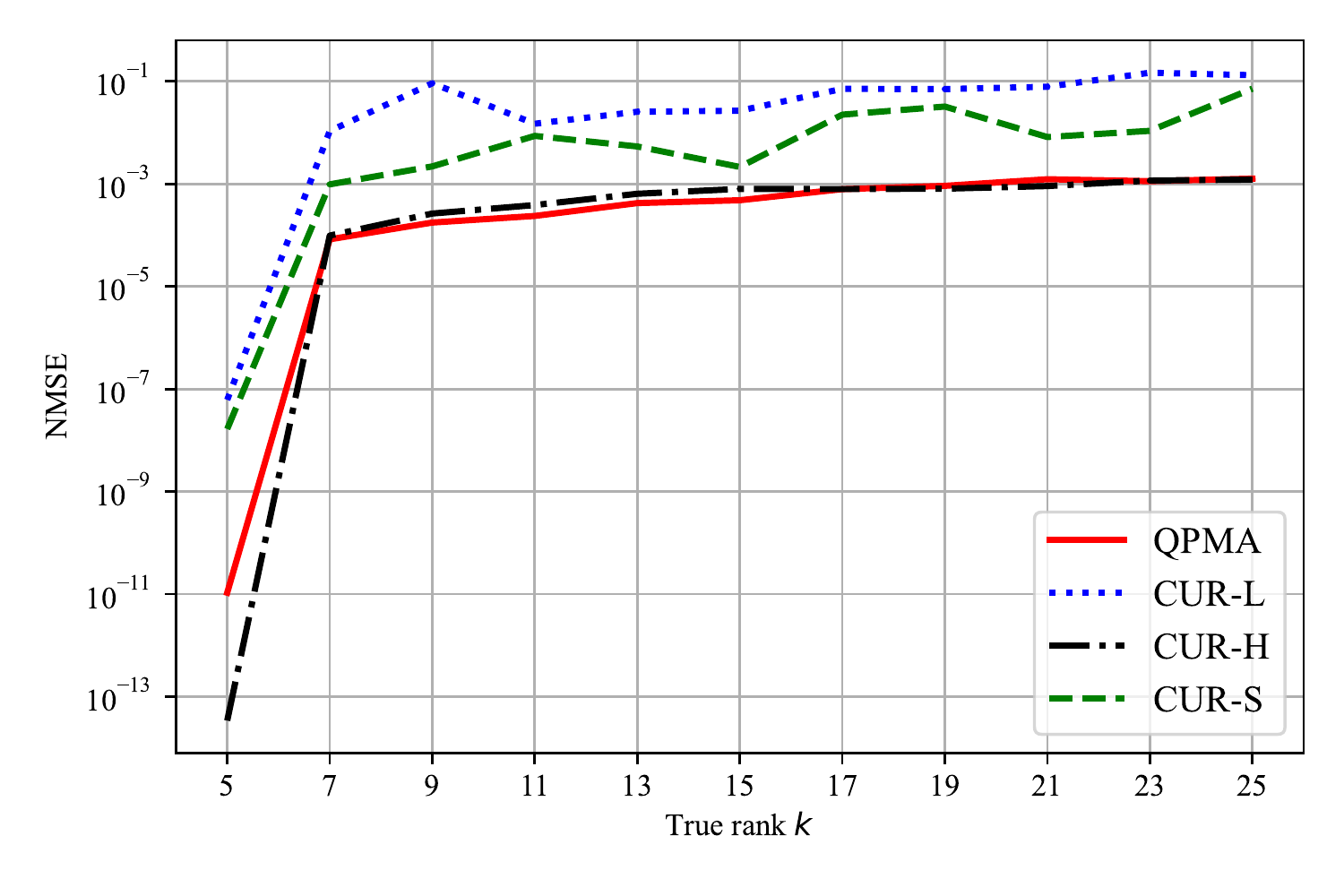}
    \vspace*{-0.3in}
    \caption{The NMSE versus the true rank $k$ of $\mathbf{M}$ with fixed $d=10$, $l=5$ and $r=5$.}\label{fig:p5}
\end{figure}

\subsubsection{Quantifying row equivalence of side-information}
We next attempt to answer the following question: {\em how much (row space) information  is being captured by the quasi-polynomial side information}\footnote{For brevity, we only compare with CUR-H since the performance of QPMA is comparable to CUR-H across various parameter regimes.}. To this end, for both QPMA and CUR-H, we fix the number of  observed columns to $d = 5$ and numerically compute the number of {\em rows} required for  CUR-H to attain the same (fixed) numerical error as that of QPMA. Additionally, we consider two cases: the polynomial coefficient matrix, $\mathbf{Q}$ is {\bf dense} (generated as in the previous experiment), and $\mathbf{Q}$ is {\bf sparse} (generated by randomly puncturing $30\%$ of the entries). The rest of the data is generated as before, with parameters $l=\{3,5\}$ and $k=\{30,50\}$. The results for these  experiments are shown in Fig. \ref{fig:p4}. First, consider the dense $\mathbf{Q}$ case: for both values of $k$, observe that when $l = 5$, CUR-H requires roughly $8-9$ rows to match the error attained by Algorithm \ref{alg:alg} and similarly when $l=3$, CUR-H requires $5-6$ rows to match the error of QPMA. Thus, for dense $\mathbf{Q}$, CUR-H requires $\approx 2l$ rows to match the numerical performance of the proposed method. For sparse $\mathbf{Q}$, the effect is more pronounced, and {CUR-H requires roughly $2l - 3l$} rows to match the performance of QPMA.  These observations also consistent with Theorem \ref{eq:thm1}. With all other parameters fixed, making $\mathbf{Q}$ sparse, reducing $k$, and reducing $l$ each have the effect of increasing $\delta$  and hence decreasing the (bound on) the error attained by QPMA. 

\subsubsection{Varying $k$}
Next, we analyze the effect of varying the true rank, $k$. We generate the data exactly as done in the first experiment with $l=5$, $d=10$. We set the target rank $r = 5$ for all algorithms. The results are provided in Fig.~\ref{fig:p5}. Notice that when $k = l$, both QPMA and CUR-H are able to obtain near-perfect estimates of the true matrix with just $d = 2r = 10$ columns. As expected, the error increases with increasing $k$ (since the number of observed columns and the polynomial degree is fixed), but {saturates after $k \approx 3r$.} Again, this observation is consistent with Theorem \ref{eq:thm1}, as the error is dominated by the noise term, i.e., terms related to $\mbox{SNR}$ rather than the approximation error {$\|\mathbf{M} - \mathbf{M}_r\|_{2}^{2}$.  }

\begin{figure}[t!]
    \centering
    \includegraphics[width=9cm]{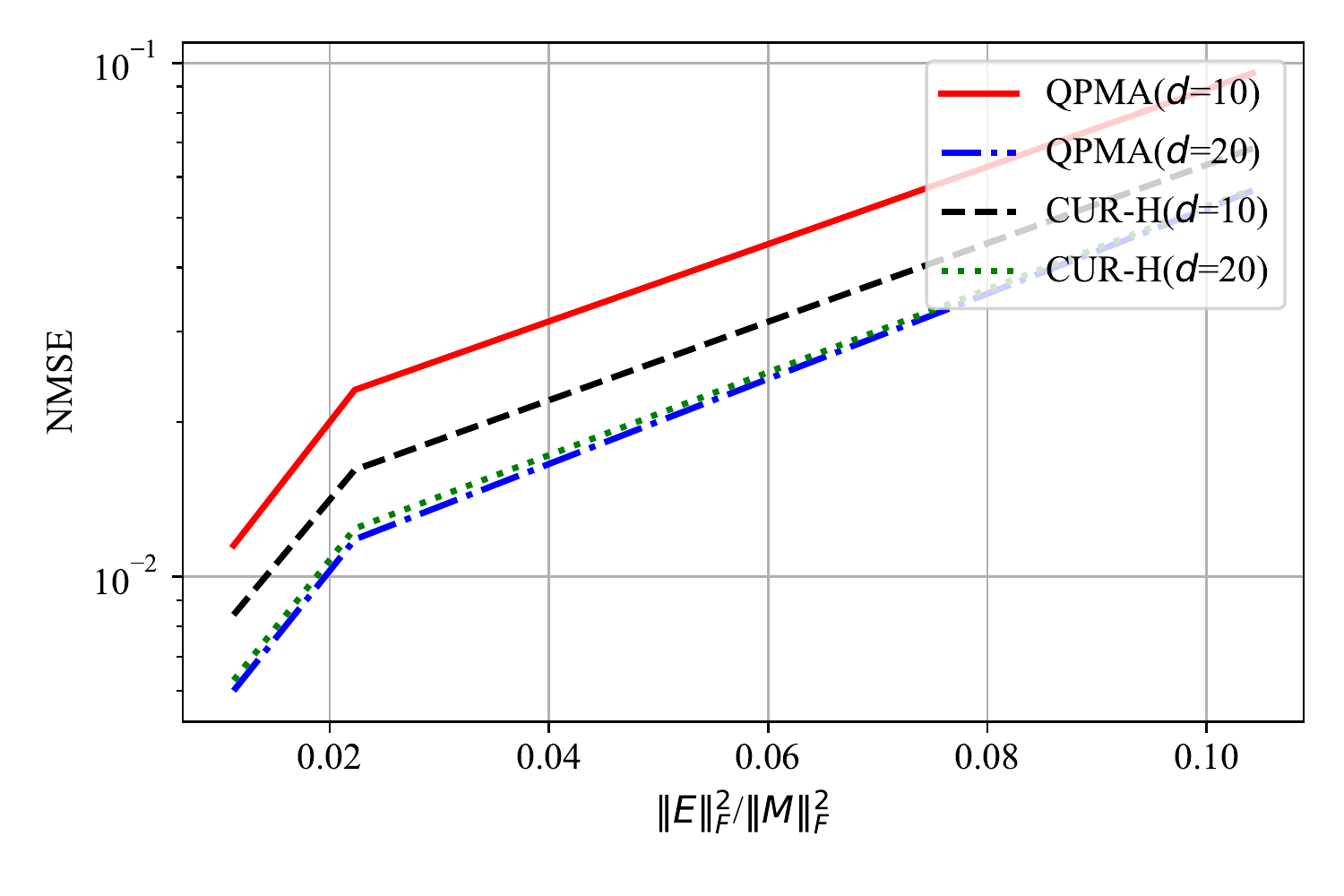}
      \vspace*{-0.3in}
    \caption{The sensitivity of QPMA to the perturbation of $\mathbf{M}$ when $d=10$ and $d=20$. The parameters are set by $r=5$, $l=5$ and $k=100$.}
       \vspace*{-0.1in}
    \label{fig:perform3}
\end{figure}

\subsubsection{Sensitivity to Noise}
We next investigate the sensitivity of QPMA to additional noise. We generate the data as done previously with $k = 100$, $d = \{10, 20\}$, and vary the noise standard deviation, $\sigma = \{10^{-4},10^{-3},10^{-2},10^{-1}\}$. We provide the results in Fig. \ref{fig:perform3}. As expected, the performance of both QPMA and CUR-H degrades as the noise increases. {Furthermore, observe that for $d = 10$, CUR-H is more robust to noise. We believe that this is because in the regime of low $d$, the unrecoverable energy term of Theorem \ref{eq:thm1} dominates, while CUR-H has a lower effective bound due to observing significantly more samples than QPMA. For $d=20$, we notice that QPMA is at least as robust as CUR-H and this is in accordnace with Theorem \ref{eq:thm1} as well, as in this regime error is dominated by the imperfect side-information (large $\mathbf{E}$) terms.}


\subsection{Real data} 
We evaluate QPMA on the real Hessian eigenvalues matrix of a chemical system provided in \cite{quiton2022toward}. In particular, we consider the $CF_3CH_3$ reaction system. For this system, the true matrix, $\mathbf{M} \in \RR^{24 \times 52}$, we observed that there is a good singular value separation at $r = 5$ and more specifically, $\sigma_{1}(\mathbf{M})=4.857$, $\sigma_{6}(\mathbf{M})=5.401\times10^{-3}$ and the matrix is full rank, i.e., $k=24$. 
Informed by the singular value gap, we chose the target rank  $r = 5$. Based on the methodology proposed in \cite{quiton2022toward}, we selected $ l =5$ and $s=[1+0.01*[m]]$. For more detailed data description, see \cite{quiton2022toward}. 

\begin{figure}[t!]
    \centering
    \includegraphics[width=9.15cm]{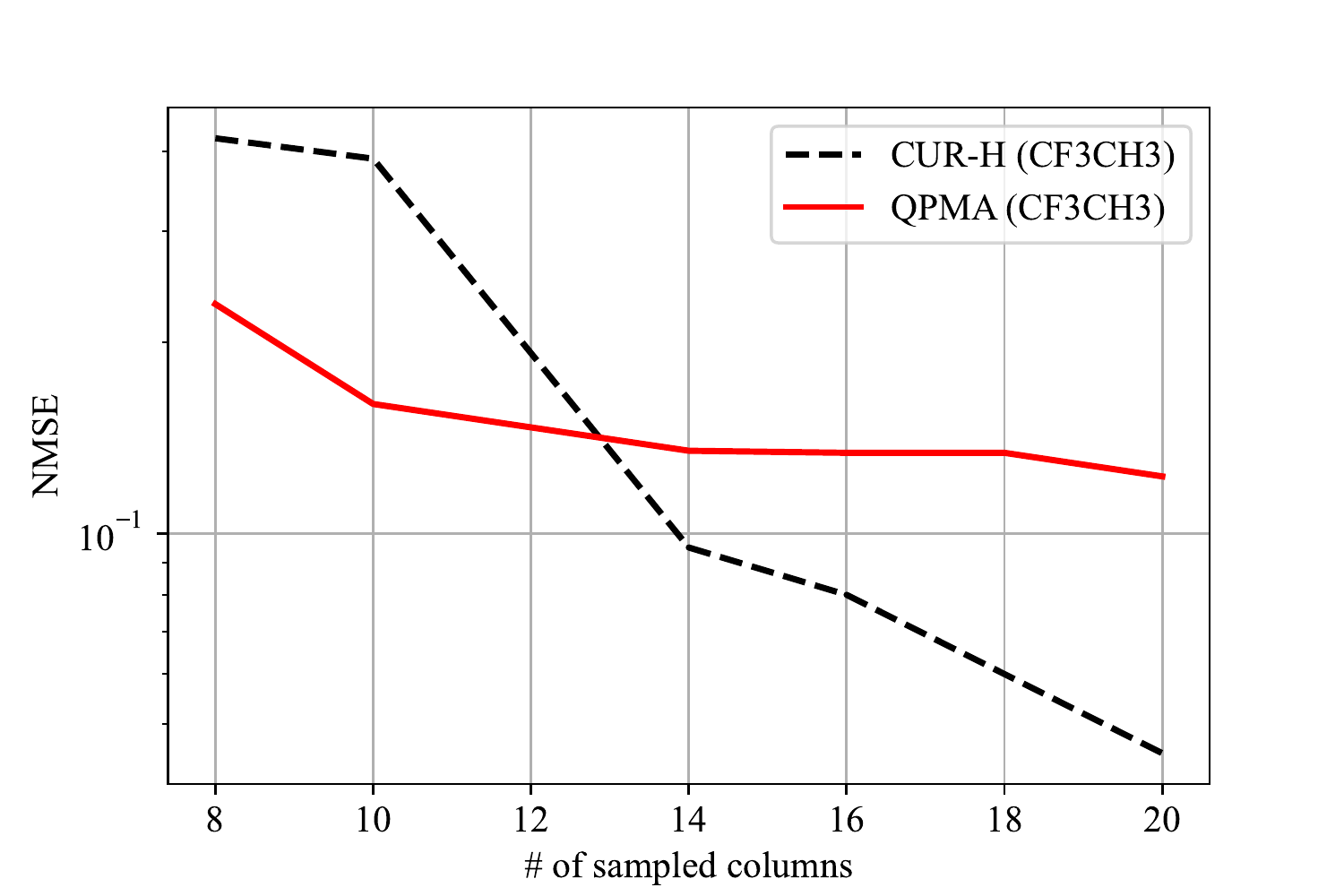}
    \vspace*{-0.3in}
    \caption{The NMSE versus the number of sampled columns $d$ for CF3CH3 chemical system with $l=5$, $r=5$ and $k=24$.}
    \label{fig:perform7}
\end{figure}

{To simulate the setting of column-sampling only and limit the access to true rows, in the implementation of CUR-H, we provide $d$ estimated rows (via QPMA) and $24-d$ true rows.} We implemented QPMA with $\eta =0.01$ and $ T = 1500$. We present the results in Fig.~\ref{fig:perform7}. We observe that in the low-sample regime ($d \leq 13$), QPMA outperforms CUR-H indicating that the side information is effectively being exploited by QPMA, whereas in the large sample regime, CUR-H tends to perform better than QPMA. However, we emphasize that (a) sampling more columns is often prohibitively more expensive in practice, and (b) CUR-H cannot be implemented in reality, since in general, one does not have access to the row information.  Thus, we see that the proposed algorithm does in fact work well for the application that motivated the our algorithm.  QPMA provides a tool by which the computation  of key quantities for VTST can be reduced while offering good approximation performance.  Our theoretical analysis provides strategies by which to understand VTST from a signal processing perspective.

\section{Conclusions}
In this paper, we formulated a novel matrix approximation problem wherein we observe are a few arbitrary columns of a high-rank matrix. In order to make the problem tractable, and inspired by problems in quantum chemistry, we imposed a quasi-polynomial structural information. We designed and analyzed an algorithm dubbed Quasi-Polynomial Matrix Approximation (QPMA) to solve the above problem and derived theoretical guarantees. { Our main guarantees show that the results are only slightly worse than  state-of-the-art results in matrix approximation, albeit this work considers a significantly harder problem}. Finally, we also provided several numerical experiments that validate our main guarantees. Specifically, we showed that (i) in the low-sample regime, the proposed method is roughly two-three orders of magnitude better than CUR+ \cite{xu2015cur}; (ii) in general, the polynomial structural information with degree $l$ is roughly equivalent to observing $2l-3l$ rows of the original matrix; and (iii) choosing the appropriate target rank is critical due to the sensitivity of the matrix approximation strategies to rank mismatch. Via simulation, it is shown that the error saturates after $k \approx 3r$. Finally, we show that our proposed methods work for the motivating quantum chemistry problem. We propose to characterize the classes of side information $\mathbf{S}$ that our approach can handle in future work.


\section*{Acknowledgements}
This work is funded in part by one or more of the following grants: NSF CCF-1817200, ARO W911NF1910269, DOE DE-SC0021417,
Swedish Research Council 2018-04359, NSF CCF-2008927, NSF CCF-2200221, ONR 503400-78050, ONR N00014-15-1-2550 and USC+Amazon Trust AI center.

\appendix
\section{Appendix}

Here, we prove Lemmas~\ref{eq:lem1}, \ref{eq:lem3} and \ref{eq:lem2} and complete the proof of Theorem~\ref{eq:thm1}.
Throughout the proof, we invoke the following norm property. For a matrix $\mathbf{A}$, the ${\Vert\cdot\Vert}_{2}$ norm is given as,
{\begin{align}
{\Vert \mathbf{A} \Vert}_2 = \max_{\Vert \mathbf{x}_{2}\Vert =1}\Vert \mathbf{Ax} \Vert_{2} = \max_{\Vert \mathbf{x}\Vert\neq0} \frac{\Vert\mathbf{Ax} \Vert_{2}}{\Vert \mathbf{x} \Vert_{2}}=\sigma_{1}(\mathbf{A}).\nonumber
\end{align}} 
For a projection matrix, $\mathbf{P}_{\uA}$ it is easy to see that $\sigma_{1}\left(\mathbf{P}_{\uA}\right)=1$.

\subsection{Proof of Lemma~\ref{eq:lem1}}\label{sec:appendix_a} 
We use Lemma~\ref{eq:lem4} and Lemma~\ref{eq:lem5} to prove Lemma~\ref{eq:lem1}. Lemma~\ref{eq:lem4} was proved in \cite{xu2015cur}.

{{
\begin{lemma} \cite{xu2015cur}\label{eq:lem4} With probability $1-c_1r^{-10}$, and if $d \geq c_2\mu r\ln r$, for constants $c_{1}>0$ and $c_{2}>0$, we have
\begin{align}
    {\left\Vert \mathbf{M}-\mathbf{P}_{\uA}\mathbf{M}\right\Vert}_{2}^{2} \leq \sigma^{2}_{r+1}\left(\mathbf{M}\right)\left(1+2\frac{n}{d}\right).\nonumber
\end{align}
\end{lemma}

{
\begin{lemma}\label{eq:lem5} Assume that there exists a $\delta_1>0$ that satisfies 
Assumption~\ref{as:as1}. Then, if $d \geq c\mu r\ln r$, we have,
\begin{align}
&{\left\Vert\mathbf{P}_{\uA}\mathbf{M}-\mathbf{P}_{\uA}\mathbf{M}\mathbf{P}_{\vqsh}\right\Vert_{2}^{2}}\nonumber\\ &\;\;\leq 2 \sigma_{r+1}^{2}(\mathbf{M})+16\sigma_{1}^{2}(\mathbf{M})\left\Vert\mathbf{E}\right\Vert_{F}^{2}\left(\frac{1}{\delta_1^2}+ \frac{1}{\delta_s^2}\right)\nonumber,
\end{align} where $\delta_s := \frac{\left\Vert\left(\mathbf{S\Psi}\right)^{\dagger}\mathbf{S}\right\Vert_{F}}{\sigma_{l}(\mathbf{QS})}$ and $c>0$ is a constant. 
\end{lemma}}}}

\begin{proof}[Proof of Lemma \ref{eq:lem5}] 
We first define some preliminaries that are required to prove Lemma \ref{eq:lem5}. We use the following definition of Canonical angles as a distance measure between subspaces. 

\begin{definition}[Canonical angle between subspaces \cite{stewart1998perturbation}]\label{def:cano_angle}
Let $\mathbf{X} \in \RR^{n \times  k}$ and $\mathbf{Y} \in \RR^{n \times k}$ be matrices, whose columns form orthonormal basis for column space of each. Let $\gamma_1 \geq \dots \geq \gamma_k$ be the singular values of $\mathbf{X}^{\trasp}\mathbf{Y}$. Then, the canonical angles between the column subspace of $\mathbf{X}$ and $\mathbf{Y}$ are defined as
\begin{align}
    \theta_{i}=\cos^{-1}\gamma_i , \; i \in [k].\nonumber
\end{align} 
\end{definition}

Next, we introduce Wedin's theorem \cite{stewart1998perturbation} that is used to bound the distance between the subspaces of two matrices. For this part, consider $\mathbf{X} \in \mathbb{R}^{n\times m}$ with rank $k$ and let $\tilde{\mathbf{X}}={\mathbf{X}}+{\mathbf{E}}$ be a perturbation of $\mathbf{X}$. Denote the SVDs of $\mathbf{X}, \mathbf{\tilde{X}}$ as 

\begin{align}\label{eq:svd_X}
    \mathbf{X}&\stackrel{\text{SVD}}{=}\mathbf{U}_{1}\mathbf{\Sigma}_{1}\mathbf{V}_{1}^\transp + \mathbf{U}_{2}\mathbf{\Sigma}_{2}\mathbf{V}_{2}^\transp,
\end{align} 
and 
\begin{align}\label{eq:svd_X_p}
    \tilde{\mathbf{X}}&\stackrel{\text{SVD}}{=}
     \tilde{\mathbf{U}}_{1}\tilde{\mathbf{\Sigma}}_{1}\tilde{\mathbf{V}}_{1}^\transp + \tilde{\mathbf{U}}_{2}\tilde{\mathbf{\Sigma}}_{2}\tilde{\mathbf{V}}_{2}^\transp.
\end{align}
where the singular values are not necessarily presented in a descending order. Then, Wedin's theorem says the following. 

\begin{theorem}
[{{\bf Wedin's theorem} \cite{stewart1998perturbation}}] \label{def:wedin}
Let $\mathbf{\Phi}$ denote the matrix of canonical angles between $\mathbf{U}_1$ and $\mathbf{ \tilde{U}}_1$, and  $\mathbf{\Theta}$ be the matrix of canonical angles between $\mathbf{V}_1$ and $ \tilde{\mathbf{V}}_1$ in \eqref{eq:svd_X} and \eqref{eq:svd_X_p} respectively. And, if there is a $\delta > 0$ such that 
\begin{align}
    \delta=\min\{\min_{1\leq i\leq k, k \leq j \leq m}\vert\sigma_{i}(\mathbf{X})-\sigma_{j}(\mathbf{\tilde{\mathbf{X}}})\vert,\min_{1\leq i \leq k} \sigma_{i}(\mathbf{X})\}\nonumber
\end{align} then 
\begin{align}
    \sqrt{{\left\Vert\sin \mathbf{\Theta}\right\Vert}_{F}^{2} + {\left\Vert\sin \mathbf{\Phi}\right\Vert}_{F}^{2}} \leq \frac{\sqrt{{\left\Vert \mathbf{R}_1 \right\Vert}_{F}^{2}+{\left\Vert \mathbf{S}_1 \right\Vert}_{F}^{2}}}{\delta},\nonumber
\end{align} 
where $\mathbf{R}_1$, ``residual between the column spaces'' is
\begin{align}
    \mathbf{R}_1 = \mathbf{X}\tilde{\mathbf{ V}}_1-\tilde{\mathbf{U}}_1{\tilde{\mathbf{ \Sigma}}_1}\nonumber 
\end{align} 
and $\mathbf{S}_1$, the ``residual between the row spaces'' is
\begin{align}
    \mathbf{S}_1 = \mathbf{ X}^{\trasp}{\tilde{\mathbf{ U}}_1}-{\tilde{\mathbf{V}}_1}{\tilde{\mathbf{\Sigma}}_1}.\nonumber
\end{align} 
\end{theorem}

We also use the following theorem that discusses the connection between canonical angles and projections.
\begin{theorem}{({\bf The connection between canonical angles and projections} \cite{stewart1990matrix})}\label{def:thm3} Let $\mathbf{P}_{{\mathbf{V}_{1}}}$ and $\mathbf{P}_{\tilde{\mathbf{V}}_{1}}$ denote the orthogonal projections onto $\mathbf{V}_{1}$ and $\tilde{\mathbf{V}}_{1}$ respectively. Let $\mathbf{\Theta}$ be the matrix of canonical angles between $\mathbf{V}_{1}$ and $\tilde{\mathbf{V}}_{1}$. Define $\Vert \mathbf{P}_{\mathbf{V}_{1}}-\mathbf{P}_{\tilde{\mathbf{V}}_{1}} \Vert_{F}$. Then, 
\begin{align}
    \left\Vert \mathbf{P}_{\mathbf{V}_{1}}-\mathbf{P}_{\tilde{\mathbf{V}}_{1}} \right\Vert_{F}=\sqrt{2}\left\Vert\sin \mathbf{\Theta}\right\Vert_{F}.
\end{align}
\end{theorem}

{ Now, the proof of Lemma \ref{eq:lem5} follows from two applications of Theorem \ref{def:wedin} and Theorem \ref{def:thm3}. For the first application, we invoke Theorem \ref{def:wedin} with $\mathbf{X} \equiv \mathbf{QS}, \mathbf{\tilde{X}} \equiv \mathbf{\hat{Q}S}$. Recall that $\mathbf{Q}$ indicates the true polynomial coefficient matrix from \eqref{eq:structure_M} and $\mathbf{\hat{Q}}$ is the estimate that is obtained from \eqref{eq:update_q}. Then, we have 
\begin{align}\label{eq:delta_2}
    \delta_2&=\min\{\min_{1\leq i\leq l, l+1 \leq j \leq m}\vert\sigma_{i}(\mathbf{QS})-\sigma_{j}(\mathbf{\hat{Q}S})\vert,\min_{1\leq i \leq l} \sigma_{i}(\mathbf{QS})\}\nonumber\\&\stackrel{(a)}{=} \sigma_{l}(\mathbf{QS}),
\end{align}  
where (a) follows from using the fact that $\mathrm{rank}(\mathbf{\hat{Q}S}) \leq l$ and thus $\sigma_{j}(\mathbf{\hat{Q}S})=0$ for $j > l$. Therefore, $\delta_2=\sigma_{l}(\mathbf{QS})$. Next, we compute the residuals required for Wedin's theorem as follows
\begin{align}\label{eq:res_t}
    \mathbf{T} := \mathbf{\hat{Q}S}\mathbf{V}_{QS} - \mathbf{U}_{QS}\mathbf{\Sigma}_{QS}
\end{align} 
and 
\begin{align}\label{eq:res_w}
    \mathbf {W}:=\mathbf{(\hat{Q}S)}^{\trasp}\mathbf{U}_{QS} - \mathbf{V}_{QS}\mathbf{\Sigma}_{QS}.
\end{align} 

Let 
\begin{align}\label{eq:d}
  \mathbf{D}:=\mathbf{\hat{Q}S}-\mathbf{QS}.
\end{align}

Then, we invoke Theorem \ref{def:thm3} with $\mathbf{V}_1 \equiv \mathbf{V}_{QS}$ and $\mathbf{\tilde{V}}_1 \equiv \mathbf{\hat{V}}_{QS}$ to obtain 
\begin{align}\label{eq:p_v_qs}
{\left\Vert\mathbf{P}_{\mathbf{V}_{QS}}-\mathbf{P}_{\vqsh}\right\Vert}_{2}^{2}&={\left\Vert\mathbf{V}_{QS}\mathbf{V}_{QS}^{\trasp}-\vqsh\vqsh^{\trasp}\right\Vert}_{2}^{2}\nonumber\\
&\stackrel{(a)}{\leq}{\left\Vert\mathbf{V}_{QS}\mathbf{V}_{QS}^{\trasp}-\vqsh\vqsh^{\trasp}\right\Vert}_{F}^{2}\nonumber\\
&=2{\Vert \sin\mathbf{\Theta_1} \Vert}_{F}^{2}\nonumber\\
&\leq 2\left(\frac{{\Vert \mathbf{T} \Vert}_{F}^{2}+{\Vert \mathbf{W} \Vert}_{F}^{2}}{\delta_2^2}-m\right)\nonumber\\
&\leq 2\left(\frac{{\Vert \mathbf{T} \Vert}_{F}^{2}+{\Vert \mathbf{W} \Vert}_{F}^{2}}{\delta_2^2}\right)\nonumber\\
&\stackrel{(b)}{\leq} \frac{4{\Vert \mathbf{D} \Vert}_{F}^{2}}{\delta_2^2} = \frac{4{\Vert \mathbf{D} \Vert}_{F}^{2}}{\sigma_{l}^2(\mathbf{QS})},
\end{align} where (a) follows from using $\Vert \cdot \Vert_{2}^{2} \leq \Vert \cdot \Vert_{F}^{2}$ and (b) is due to 
\begin{align}\label{eq:bound_R}
    {\Vert \mathbf{T} \Vert}_{F}^{2}={\Vert \mathbf{\hat{Q}S}\mathbf{V}_{QS} - \mathbf{U}_{QS}\mathbf{\Sigma}_{QS}\Vert}_{F}^{2} \leq {\Vert \mathbf{D}\mathbf{V}_{QS} \Vert}_{F}^{2}
    \leq {\Vert \mathbf{D} \Vert}_{F}^{2},  
\end{align} 
with a similar bound for $\mathbf{W}$. Now, we further bound ${\Vert\mathbf{D}\Vert}_{F}^{2}$ in \eqref{eq:d}. Given $\mathbf{A}$, $\mathbf{S}$ and $\mathbf{\Psi}$,  $\hat{\mathbf{Q}}$ is obtained by solving the unconditioned least-squares problem \eqref{eq:opt_Q}. This problem can be solved analytically. Since the rows of $\mathbf{S\Psi}$ are independent, the least-squares approximation problem has the unique solution $\hat{\mathbf{Q}}=\mathbf{A}\left(\mathbf{S\Psi}\right)^{\trasp}\left(\mathbf{S\Psi}\left(\mathbf{S\Psi}\right)^{\trasp}\right)^{-1}$ \cite[p.155]{Rencher2012methods}. Therefore, we have the following bound for ${\Vert\mathbf{D}\Vert}_{F}^{2}$,

\begin{align}\label{eq:d2}
    {\Vert\mathbf{D}\Vert}_{F}^{2}&\;={\Vert\hat{\mathbf{Q}}\mathbf{S}-\mathbf{QS}\Vert}_{F}^{2}\nonumber\\
    &\;= \left\Vert \mathbf{A}\left(\mathbf{S\Psi}\right)^{\trasp}\left(\mathbf{S\Psi}\left(\mathbf{S\Psi}\right)^{\trasp}\right)^{-1}\mathbf{S} -\mathbf{QS}\;\right\Vert_{F}^{2}\nonumber\\
    &\; = \left\Vert \mathbf{M\Psi}\left(\mathbf{S\Psi}\right)^{\trasp}\left(\mathbf{S\Psi}\left(\mathbf{S\Psi}\right)^{\trasp}\right)^{-1}\mathbf{S} -\mathbf{QS}\;\right\Vert_{F}^{2}\nonumber\\
    &\; = \left\Vert \mathbf{(QS+E)\Psi}\left(\mathbf{S\Psi}\right)^{\trasp}\left(\mathbf{S\Psi}\left(\mathbf{S\Psi}\right)^{\trasp}\right)^{-1}\mathbf{S} -\mathbf{QS}\;\right\Vert_{F}^{2}\nonumber\\
    &\;= \Vert \mathbf{QS\Psi}\left(\mathbf{S\Psi}\right)^{\trasp}\left(\mathbf{S\Psi}\left(\mathbf{S\Psi}\right)^{\trasp}\right)^{-1}\mathbf{S}\nonumber \\ &\;\;\;\;\;\;\;\;\;\;\;\;\; \;\;\;+\mathbf{E\Psi}\left(\mathbf{S\Psi}\right)^{\trasp}\left(\mathbf{S\Psi}\left(\mathbf{S\Psi}\right)^{\trasp}\right)^{-1}\mathbf{S} -\mathbf{QS}\;\Vert_{F}^{2}\nonumber\\
    &\;= \left\Vert\mathbf{E\Psi}\left(\mathbf{S\Psi}\right)^{\trasp}\left(\mathbf{S\Psi}\left(\mathbf{S\Psi}\right)^{\trasp}\right)^{-1}\mathbf{S} \right\Vert_{F}^{2}\nonumber\\
    &\;= \left\Vert\mathbf{E\Psi}\left(\mathbf{S\Psi}\right)^{\dagger}\mathbf{S} \right\Vert_{F}^{2}\nonumber\\
    &\; \stackrel{(a)}{\leq} \left\Vert\mathbf{E}\right\Vert_{F}^{2}\left\Vert\left(\mathbf{S\Psi}\right)^{\dagger}\mathbf{S}\right\Vert_{F}^{2}, 
\end{align} where (a) is due to the matrix norm inequality.

Thus, using \eqref{eq:d2} and \eqref{eq:p_v_qs}, we have, 
\begin{align}\label{eq:p_v_qs3}
    {\left\Vert\mathbf{P}_{\mathbf{V}_{QS}}-\mathbf{P}_{\vqsh}\right\Vert}_{2}^{2} \leq \frac{4\left\Vert\mathbf{E}\right\Vert_{F}^{2}}{\sigma_{l}^2(\mathbf{QS})}\left\Vert\left(\mathbf{S\Psi}\right)^{\dagger}\mathbf{S}\right\Vert_{F}^{2}.
\end{align}
}

Next, using a similar approach, we apply Theorems \ref{def:wedin} and \ref{def:thm3} with $\mathbf{X} \equiv \mathbf{M}$ and $\mathbf{\tilde{X}} \equiv \mathbf{QS}$. 

By plugging \eqref{eq:p_v_qs3} and \eqref{eq:p_v_m_qs2} into \eqref{eq:p_v_m_qs}, we have
\begin{align}
{\left\Vert\mathbf{P}_{\mathbf{V}_{M}}-\mathbf{P}_{\vqsh}\right\Vert}_{2}^{2} \leq 8\left\Vert\mathbf{E}\right\Vert_{F}^{2}\left(\frac{1}{\delta_1^2} + \frac{\left\Vert\left(\mathbf{S\Psi}\right)^{\dagger}\mathbf{S}\right\Vert_{F}^{2}}{\sigma_{l}^2(\mathbf{QS})}\right).
\end{align}
First, we demonstrate the minimum eigengap separation condition as follows. Let
\begin{align}
    \delta_1=\min\{\min_{1\leq i\leq l, l+1 \leq j \leq m}\vert\sigma_{i}(\mathbf{QS})-\sigma_{j}(\mathbf{M})\vert,\min_{1\leq i \leq l} \sigma_{i}(\mathbf{QS})\}\nonumber
\end{align} 

Notice that since $\mbox{rank}(\mathbf{M}) = k \gg r$ and $\mbox{rank}(\mathbf{QS}) = l \geq r$, the first term above attains the minimum and thus $\delta_1 = \sigma_r(\mathbf{M - E}) - \sigma_{r+1}(\mathbf{M})$. This is bounded away from zero owing to Assumption \ref{as:as1}. We next compute the residuals as follows 
\begin{align}\label{eq:res_r}
    \mathbf{R} := \mathbf{QS}\vM - \uM\sM
\end{align} and $\mathbf{S}$ is defined as the residual between the row space $\vM$ and $\vqs$ as
\begin{align}\label{eq:res_s}
    \mathbf {S}:=\mathbf{ (QS)}^{\trasp}\uM - \vM\sM.
\end{align}

Then,
\begin{align}\label{eq:p_v_m_qs2}
{\left\Vert\mathbf{P}_{\mathbf{V}_{M}}-\mathbf{P}_{\mathbf{V}_{QS}}\right\Vert}_{2}^{2}&=\left\Vert\mathbf{V}_{M}\mathbf{V}_{M}^{\trasp}-\mathbf{V}_{QS}\mathbf{V}_{QS}^{\trasp}\right\Vert_{2}^{2}\nonumber\\
&\stackrel{(a)}{\leq}\left\Vert\mathbf{V}_{M}\mathbf{V}_{M}^{\trasp}-\mathbf{V}_{QS}\mathbf{V}_{QS}^{\trasp}\right\Vert_{F}^{2}\nonumber\\
&\stackrel{(b)}{=}2{\Vert \sin\mathbf{\Theta} \Vert}_{F}^{2}\nonumber\\
&\stackrel{(c)}{\leq} 2\left(\frac{{\Vert \mathbf{R} \Vert}_{F}^{2}+{\Vert \mathbf{S} \Vert}_{F}^{2}}{\delta_1^2}-m\right)\nonumber\\
&\leq 2\left(\frac{{\Vert \mathbf{R} \Vert}_{F}^{2}+{\Vert \mathbf{S} \Vert}_{F}^{2}}{\delta_1^2}\right),\nonumber\\
&\stackrel{(d)}{\leq} \frac{4{\Vert \mathbf{E} \Vert}_{F}^{2}}{\delta_1^2}
\end{align} where the inequality (a) is due to $\Vert \cdot \Vert_{2}^{2} \leq \Vert \cdot \Vert_{F}^{2}$ and (b) and (c) are from  Theorems~\ref{def:wedin} and \ref{def:thm3} and respectively. (d) is due to 
\begin{align}\label{eq:bound_R}
    {\Vert \mathbf{R} \Vert}_{F}^{2}={\Vert \mathbf{ QS}\vM-\uM\sM \Vert}_{F}^{2} \leq {\Vert \mathbf{E}\vM \Vert}_{F}^{2}
    \leq {\Vert \mathbf{E} \Vert}_{F}^{2},
\end{align} with a similar bound for $\mathbf{S}$.
With these bounds, we prove Lemma \ref{eq:lem5} as follows

\begin{align}
    &{\left\Vert\mathbf{P}_{\uA}\mathbf{M}-\mathbf{P}_{\uA}\mathbf{M}\mathbf{P}_{\vqsh}\right\Vert_{2}^{2}}\nonumber\\
    &\stackrel{(a)}{\leq} {\left\Vert  \mathbf{P}_{\uA}\right\Vert}_{2}^{2}{\left\Vert  \mathbf{M}-\mathbf{M}\mathbf{P}_{\vqsh}\right\Vert}_{2}^{2}\nonumber\\
    &={\left\Vert  \mathbf{M}-\mathbf{M}\mathbf{P}_{\vqsh}\right\Vert}_{2}^{2}\nonumber\\
    &={\left\Vert  \mathbf{M}-\mathbf{M}\mathbf{P}_{\mathbf{V}_{M}}+\mathbf{M}\mathbf{P}_{\mathbf{V}_{M}}-\mathbf{M}\mathbf{P}_{\vqsh}\right\Vert}_{2}^{2}\nonumber\\
    &\stackrel{(b)}{\leq} 2 {\left\Vert  \mathbf{M}-\mathbf{M}\mathbf{P}_{\mathbf{V}_{M}}\right\Vert}_{2}^{2}+2 {\left\Vert\mathbf{M}\mathbf{P}_{\mathbf{V}_{M}}-\mathbf{M}\mathbf{P}_{\vqsh}\right\Vert}_{2}^{2}\nonumber\\
    &\stackrel{(c)}{\leq} 
    2 \sigma_{r+1}^{2}(\mathbf{M})+2{\left\Vert\mathbf{M}\mathbf{P}_{\mathbf{V}_{M}}-\mathbf{M}\mathbf{P}_{\vqsh}\right\Vert}_{2}^{2}\nonumber\\
    &\stackrel{(d)}{\leq}2 \sigma_{r+1}^{2}(\mathbf{M})+2{\left\Vert\mathbf{M}\right\Vert}_{2}^{2}{\left\Vert\mathbf{P}_{\mathbf{V}_{M}}-\mathbf{P}_{\vqsh}\right\Vert}_{2}^{2}\nonumber\\
&\stackrel{(e)}{\leq}2 \sigma_{r+1}^{2}(\mathbf{M})+2\sigma_{1}^{2}(\mathbf{M}){\left\Vert\mathbf{P}_{\mathbf{V}_{M}}-\mathbf{P}_{\vqsh}\right\Vert}_{2}^{2}\label{eq:a},
\end{align} where the inequality (a) and (d) is due to matrix norm inequality and (b) is due to the fact that for $a,b \geq 0, (a+b)^2 \leq 2(a^2+b^2)$. Inequalities (c) and (e) are derived from the operator norm property. {{Next, we bound ${\left\Vert\mathbf{P}_{\mathbf{V}_{M}}-\mathbf{P}_{\vqsh}\right\Vert}_{2}^{2}$.
\begin{align}
&{\left\Vert\mathbf{P}_{\mathbf{V}_{M}}-\mathbf{P}_{\vqsh}\right\Vert}_{2}^{2}\nonumber\\
&={\left\Vert\mathbf{P}_{\mathbf{V}_{M}}-\mathbf{P}_{\mathbf{V}_{QS}}+\mathbf{P}_{\mathbf{V}_{QS}}-\mathbf{P}_{\vqsh}\right\Vert}_{2}^{2}\nonumber\\
&\stackrel{(a)}{\leq} 2{\left\Vert\mathbf{P}_{\mathbf{V}_{M}}-\mathbf{P}_{\mathbf{V}_{QS}}\right\Vert}_{2}^{2} + 2{\left\Vert\mathbf{P}_{\mathbf{V}_{QS}}-\mathbf{P}_{\vqsh}\right\Vert}_{2}^{2}\label{eq:p_v_m_qs},  
\end{align} where (a) is due to the fact that for $a,b \geq 0, (a+b)^2 \leq 2(a^2+b^2)$.

 Finally, combining everything we have
\begin{align}
    &{\left\Vert\mathbf{P}_{\uA}\mathbf{M}-\mathbf{P}_{\uA}\mathbf{M}\mathbf{P}_{\vqsh}\right\Vert_{2}^{2}} \nonumber\\
    &\leq 2 \sigma_{r+1}^{2}(\mathbf{M})+16\sigma_{1}^{2}(\mathbf{M})\left\Vert\mathbf{E}\right\Vert_{F}^{2}\left(\frac{1}{\delta_1^2}+ \frac{\left\Vert\left(\mathbf{S\Psi}\right)^{\dagger}\mathbf{S}\right\Vert_{F}^{2}}{\sigma_{l}^2(\mathbf{QS})}\right)
\end{align}}}
\end{proof}


{
Finally, we use Lemma~\ref{eq:lem4} and Lemma~\ref{eq:lem5} to prove Lemma~\ref{eq:lem1} as follows.
\begin{align}
    &\left\Vert\mathbf{M}-\mathbf{P}_{\uA}\mathbf{M}\mathbf{P}_{\vqsh}\right\Vert_{2}^{2}\;\nonumber\\
    &\;\stackrel{\text{(a)}}{\leq}2\left\Vert\mathbf{M}-\mathbf{P}_{\uA}\mathbf{M}\right\Vert_{2}^{2} + 2\left\Vert\mathbf{P}_{\uA}\mathbf{M}-\mathbf{P}_{\uA}\mathbf{M}\mathbf{P}_{\vqsh}\right\Vert_{2}^{2}\nonumber \\
    &\; \stackrel{(b)}{\leq} 2\sigma^{2}_{r+1}\left(\mathbf{M}\right)\left(3+\frac{n}{d}\right)\nonumber\\
    &\;\;\;\;\;\;\;\;\;\;\;\;\;\;+32\sigma_{1}^{2}(\mathbf{M})\left\Vert\mathbf{E}\right\Vert_{F}^{2}\left(\frac{1}{\delta_1^2}+ \frac{\left\Vert\left(\mathbf{S\Psi}\right)^{\dagger}\mathbf{S}\right\Vert_{F}^{2}}{\sigma_{l}^2(\mathbf{QS})}\right)
\end{align} The inequalities (a) is due to the fact that $(a+b)^2 \leq 2(a^2 + b^2)$ for $ a,b \geq 0$. (b) follows from Lemma~\ref{eq:lem4} and Lemma~\ref{eq:lem5}. This concludes the proof of Lemma~\ref{eq:lem1}. }

\vspace{0.2cm}


\subsection{Proof of Lemma~\ref{eq:lem2}}\label{sec:appendix_b}
Lemma~\ref{eq:lem2} ensures the strong convexity of the objective function $f(\mathbf{Z})$ in \eqref{eq:opt_Z_final} by restricting the curvature of the column sampling operator $\mathbf{\Psi}$\cite{negahban2012restricted,negahban2012unified}. Recall that $\mathbf{\Psi}$ is consisted of $d$ number of randomly chosen columns in $\mathbf{M}$. 
We first provide an additional necessary theorem from \cite{tropp2011improved} which describes the large-deviation behavior of specific types of matrix random variables.

\begin{theorem}\label{def:thm4}{\cite{tropp2011improved}} Let $\mathcal{X}$ be a finite set of positive semi definite matrices of dimension $k$. If there exists a constant $B < \infty$ such that 
\begin{align}
    \max_{\mathbf{X} \in \mathcal{X}} \lambda_{max}(\mathbf{X}) \leq B,\nonumber
\end{align} and, if we sample $\{\mathbf{X}_1,\dots,\mathbf{X}_p\}$ uniformly at random from $\mathcal{X}$ without replacement, with
\begin{align}
    &\mu_{max}:=p\lambda_{max}{\left(\EE[\mathbf{X}_1]\right)}\nonumber\\
    &\mu_{min}:=p\lambda_{min}{\left(\EE[\mathbf{X}_1]\right)}.\nonumber
\end{align} Then we have that, 
\begin{align}
    &P\left[\lambda_{max}\left(\sum_{i=1}^{p}\mathbf{X}_i\right) \geq (1+\rho)\mu_{max}\right]\nonumber\\
    &\leq k\cdot \mbox{exp} \frac{-\mu_{max}}{B}[(1+\rho)\ln(1+\rho)-\rho]\; \mbox{for}\; \rho \in [0,1)\nonumber\\
    &P\left[\lambda_{min}\left(\sum_{i=1}^{p}\mathbf{X}_i\right) \leq (1-\rho)\mu_{min}\right]\nonumber\\
    &\leq k \cdot \mbox{exp} \frac{-\mu_{min}}{B}[(1-\rho)\ln(1-\rho)+\rho]\; \mbox{for}\; \rho \in [0,1)\nonumber
\end{align}
\end{theorem}
Recall that Lemma~\ref{eq:lem2} provides a bound on the value of $\alpha$ in our definition of strong convexity in Definition~\ref{def:strong_convexity}. A way to prove a function is strictly convex is to show the Hessian of a function is everywhere positive definite \cite{boyd2004convex}. 
We can show the Hessian matrix is positive definite by bounding the smallest eigenvalue of the Hessian matrix as a positive value.

{{
Observe that $f(\mathbf{Z})$ can be reformulated as
\begin{align}
    f(\mathbf{Z})={\left\Vert \mathbf{A} -\uA\mathbf{Z}\vqsh^{\trasp}{\mathbf{\Psi}} \right\Vert}_{F}^{2}.\nonumber
\end{align} Now, we want to obtain
the Hessian matrix of $f (\mathbf{Z})$ and bound show that its smallest eigenvalue
is bounded away from zero. $f (\mathbf{Z})$ can also be expressed by
\begin{align}
    f(\mathbf{Z})&= {\left\Vert \mathbf{A} -\left(\uA\mathbf{Z}\vqsh^{\trasp}{\mathbf{\Psi}}\right) \right\Vert}_{F}^{2}\nonumber\\
    &={\left\Vert \sum_{j\in \mathcal{C}}\mathbf{m}_j -\left(\uA\mathbf{Z}\vqsh^{\trasp}\right)_j \right\Vert}_{F}^{2}\nonumber\\
    &\stackrel{(a)}{=}\sum_{j\in \mathcal{C}} {\left\Vert \mathbf{m}_j -\left(\uA\mathbf{Z}\vqsh^{\trasp}\right)_j \right\Vert}_{F}^{2},\nonumber\\
\end{align}  where (a) is because $j$ is sampled uniformly at random and  $\mathbf{m}_j$ indicates $j$-th column of $\mathbf{M}$, and by letting $\hat{\mathbf{m}}_j$ indicate the $j$-th column of $\hat{\mathbf{M}}$, we have,  
\begin{align}
f(\mathbf{Z})
   &=\sum_{j\in \mathcal{C}} {\left\Vert \mathbf{m}_j -\mathbf{\hat{m}}_j\right\Vert}_{F}^{2},\nonumber\\ &= \sum_{j\in \mathcal{C}} f_{j}(\mathbf{Z}),
\end{align} where $f_{j}(\mathbf{Z})={\left\Vert \mathbf{m}_j -\mathbf{\hat{m}}_j\right\Vert}_{F}^{2}={\left\Vert \mathbf{m}_j -\left(\uA\mathbf{Z}\vqsh^{\trasp}\right)_j \right\Vert}_{F}^{2}$. Then, taking the second-order derivative with respect to each element $z_{th}$ and $z_{pq}$ for $t,p \in [r]$ and $h,q \in [r]$, we have  
\begin{align}
\frac{\partial^2 f(\mathbf{Z})}{\partial z_{th}\partial z_{pq}}= \sum_{j\in\mathcal{C}}\frac{\partial^2 f_{j}(\mathbf{Z})}{\partial z_{th}\partial z_{pq}}\nonumber.
\end{align} We denote $\mathbf{H}$ and $\mathbf{H}_j$ as 
\begin{align}
\mathbf{H} := \frac{\partial^2 f(\mathbf{Z})}{\partial z_{th}\partial z_{pq}}\;\; \mbox{and} \;\; \mathbf{H}_{j} := \frac{\partial^2 f_{j}(\mathbf{Z})}{\partial z_{th}\partial z_{pq}},
\end{align} therefore, we have, 
\begin{align}
    \mathbf{H}=\sum_{j \in \mathcal{C}}\mathbf{H}_j.\label{eq:H_j}
\end{align} 
The first-order derivative of $\frac{\partial f_{j}(\mathbf{Z})}{\partial z_{th}}$ with respect to each element $z_{th}$, where $t \in [r]$ and $h \in [r]$, is given by 
\begin{align}
    \frac{\partial f_{j}(\mathbf{Z})}{\partial z_{th}} &= -2\sum_{i \in [n]}{\left({m}_{ij}-\sum_{t \in [r]}\sum_{h \in [r]}{u}_{A,it}{z}_{th}\hat{v}_{QS,hj}\right)}\nonumber\\
    &\;\;\;\;\;\;\;\;\;\;\;\;\;\;\;\;\;\;\;\;\;\;\;\;\;\;\;\;\;\;\;\;\;\;\;\;\;\;\cdot {u}_{A,it}\hat{v}_{QS,hj} \nonumber.
\end{align} 
The second-derivative of $f_j(\mathbf{Z})$ with respect to the component $z_{th}$ and $z_{pq}$ in $\mathbf{Z}$ for $t,p \in [r]$ and $h,q \in [r]$ is obtained by
\begin{align}
    \frac{\partial^2 f_{j}(\mathbf{Z})}{\partial z_{th}\partial z_{pq}} = 2\sum_{i \in [n]}{u}_{A,it}\hat{v}_{QS,hj}{u}_{A,ip}\hat{v}_{QS,qj}. \nonumber
\end{align} We let the second-order derivative of the ($i_1,j_1$)th and ($i_2,j_2$)  entry of $\mathbf{Z}$ be the ($r(i_{1}-1)+j_{1},r(i_{2}-1)+j_{2}$ entry of the Hessian matrix of $f_j(\mathbf{Z})$. Then, the Hessian $\mathbf{H}_{j}$ of $f_{j}(\mathbf{Z})$ is written as
\begin{align}
    \mathbf{H}_{j}  \in \RR^{r^2 \times r^2}=
    \begin{bmatrix}
    \frac{\partial^2 f_{j}(\mathbf{Z})}{\partial z_{11}\partial z_{11}} & \frac{\partial^2 f_{j}(\mathbf{Z})}{\partial z_{11}\partial z_{12}} & \dots & \frac{\partial^2 f_{j}(\mathbf{Z})}{\partial z_{11}\partial z_{rr}} \\
    \frac{\partial^2 f_{j}(\mathbf{Z})}{\partial z_{12}\partial z_{11}} & \frac{\partial^2 f_{j}(\mathbf{Z})}{\partial z_{12}\partial z_{12}} & \dots & \frac{\partial^2 f_{j}(\mathbf{Z})}{\partial z_{12}\partial z_{rr}} \\
    \vdots & \vdots & \vdots & \ddots  \\
    \frac{\partial^2 f_{j}(\mathbf{Z})}{\partial z_{rr}\partial z_{11}} & \frac{\partial^2 f_{j}(\mathbf{Z})}{\partial z_{rr}\partial z_{12}} & \dots & \frac{\partial^2 f_{j}(\mathbf{Z})}{\partial z_{rr}\partial z_{rr}} \nonumber.
    \end{bmatrix}
\end{align} With these preliminary derivations, the Hessian matrix of $f_{j}(\mathbf{Z})$ can be written as,
\begin{align}
    \mathbf{H}_{j} = 2\sum_{i\in[n]}\left[\mbox{vec}({\mathbf{u}}_{A,i}^{\trasp}{\hat{\mathbf{v}}}_{QS,j})\right]\left[\mbox{vec}(\mathbf{u}_{A,i}^{\trasp}\hat{\mathbf{v}}_{QS,j})\right]^{\trasp}, \nonumber
\end{align} where $\mathbf{u}_{A,i}$, for $i \in [n]$, is the $i$-th row of $\uA$ and $\hat{\mathbf{v}}_{QS,j}$ for $j \in [m]$, is the $j$-th row of $\vqsh$. By plugging this into \eqref{eq:H_j}, we obtain
\begin{align}
    \mathbf{H}&=\sum_{j\in\mathcal{C}}\mathbf{H}_{j} 
\end{align} Here, we invoke Theorem~\ref{def:thm4} to bound the smallest eigenvalue of $\mathbf{H}$. We use Theorem \ref{def:thm4} with $\mathbf{H} = \sum_j \mathbf{H}_j$ with with $\mathbf{X}_i \equiv \mathbf{H}_j$ in Theorem~\ref{def:thm4}. It follows that $p$ is $\vert\mathcal{C}\vert=d$. Note that $\mathbf{X}_{i}$ is a positive semi-definite matrix as the sum of positive semi-definite matrix is still a positive semi-definite. }} Now, let us bound the smallest eigenvalue of $\mathbf{H}$. We first want to bound the largest eigenvalue to obtain $B$. That is,
\begin{align}
    &\max_{j}\; \lambda_{max}\left(\sum_{i\in[n]}   \left[\mbox{vec}({\mathbf{u}}_{A,i}^{\trasp}\hat{\mathbf{v}}_{QS,j})\right]\left[\mbox{vec}(\mathbf{u}_{A,i}^{\trasp}\hat{\mathbf{ v}}_{QS,j})\right]^{\trasp}\right)\nonumber\\
    &\stackrel{(a)}{\leq}\max_{j}\sum_{i\in[n]}\lambda_{max}\left(\left[\mbox{vec}({\mathbf{u}}_{A,i}^{\trasp}\hat{\mathbf{v}}_{QS,j})\right]\left[\mbox{vec}(\mathbf{u}_{A,i}^{\trasp}\hat{\mathbf{ v}}_{QS,j})\right]^{\trasp}\right)\nonumber\\&\stackrel{(b)}{\leq} \max_{j}\sum_{i\in[n]} \left\Vert\mbox{vec}\left(\mathbf{u}_{A,i}^{\trasp}\hat{\mathbf{v}}_{QS,j}\right)\right\Vert_{2}^{2}\nonumber\\ 
    &\stackrel{(c)}{\leq}\max_{j}\sum_{i\in[n]}\;\left\Vert\mathbf{u}_{A,i}^{\trasp}\hat{\mathbf{v}}_{QS,j}\right\Vert_{F}^{2}\nonumber\\
    &\stackrel{(d)}{\leq} \sum_{i\in[n]}\;\Vert\mathbf{u}_{A,i}\Vert^{2}_{F}\max_{j}{\Vert\hat{\mathbf{v}}_{QS,j}\Vert}^{2}_{F}
    \stackrel{(e)}{\leq}\frac{\hat{\mu}^{2}r^2}{m}, \nonumber
\end{align}
where $(a)$ follows from Weyl's inequality \cite{Roger1991matrix},
$(b)$ follows since the argument of $\lambda_{\max}$ is the outer product of two vectors; $(c)$ follows from norm inequalities;  $(d)$ is from the Cauchy- Schwarz inequality and $(e)$ is due to the definition of the incoherence of a matrix defined in Definition~\ref{def:incoh}.
Next, we solve for $\lambda_{\min}(\EE[\mathbf{X}_{1}])$ to obtain $\mu_{\min}$ as,
\begin{align}
&\lambda_{\min}\left(\EE\left[\sum_{i\in[n]}\left[\mbox{vec}({\mathbf{u}}_{A,i}^{\trasp}{\hat{\mathbf{v}}}_{QS,j})\right]\left[\mbox{vec}(\mathbf{u}_{A,i}^{\trasp}\hat{\mathbf{v}}_{QS,j})\right]^{\trasp}\right]\right)\nonumber\\
\stackrel{(a)}{=}&\; \lambda_{\min}\left(\frac{({\uA \otimes \vqsh)}^{\trasp}\times \left(\uA\otimes \vqsh\right)}{m}\right)\nonumber\\
\stackrel{(b)}{=}&\frac{1}{m}\; \lambda_{\min}\left({(\uA \otimes \vqsh)}^{\trasp}\times \left(\uA\otimes \vqsh\right)\right)\nonumber\\
   \stackrel{(c)}{=}&\frac{1}{m},\nonumber
\end{align}

where $\otimes$ denotes the Kronecker product. The equality $(a)$ is because $j$-th row $\mathbf{\hat{v}}_{qs,j}$ for $j \in [m]$ are assumed to be randomly sampled from the rows of $\vqsh$ and (b) is due to the scalar multiplication rule for eigenvalues and $(c)$ follows since $\uA$ and $\vqsh$ have orthonormal columns. Finally, with $B=\frac{\hat{\mu}^2r^2}{m}$, $\mu_{\min}=\frac{d}{m}$ and $\rho=\frac{1}{2}$ in hand, we have, \begin{align}
    P\left\{\lambda_{\min}({\bf H})\leq \frac{d}{2m}\right\} \leq \frac{r^2}{6}e^{\frac{-d}{\hat{\mu}^{2}r^{2}}} =  e^{\frac{-d}{\hat{\mu}^{2}r^{2}}+\frac{\ln r}{3}}.\nonumber
\end{align}
This expression can be algebraically manipulated, such that 
with a probability $1-r^{-c_{1}}$, where
$c_1>0$ is a constant,
and if $ d \geq c_{2}{\hat{\mu}}^{2}r^{2}\ln r$ for a constant $c_{2}>0$, we have,
\begin{align}
    \lambda_{\min}(\mathbf{H}) \geq \frac{d}{2m}. \nonumber
\end{align}

\subsection{Proof of Lemma~\ref{eq:lem3}}\label{sec:appendix_c}
In Lemma~\ref{eq:lem3}, we bound the error between our projected true matrix and our final
estimate of the reconstructed matrix as follows. We note that
$\hat{\mathbf{M}}=\uA\hat{\mathbf{Z}}\vqsh^{\trasp}$ and hence $\hat{\mathbf{Z}}=\uA^{\trasp}\hat{\mathbf{M}}\vqsh$. Also recall that $\mathbf{Z}=\uA^{\trasp}\mathbf{M}\vqsh$.
{
\begin{align}
    &\left\Vert\mathbf{P}_{\uA}\mathbf{M}\mathbf{P}_{\vqsh}-\uA\hat{\mathbf{Z}}\vqsh^{\trasp}\right\Vert_{2}^{2}\nonumber\\
    &\;=
    \left\Vert \uA\uA^{\trasp}\mathbf{M}\vqsh\vqsh^{\trasp}-\uA\hat{\mathbf{Z}}\vqsh^{\trasp}\right\Vert_{2}^{2}\nonumber\\
    &\;\stackrel{(a)}{=}\Vert\uA\mathbf{Z}\vqsh-\uA\hat{\mathbf{Z}}\vqsh^{\trasp} \Vert_2^2\nonumber\\
    &\;\stackrel{(b)}{\leq} \left\Vert\uA\right\Vert_2^2\left\Vert\mathbf{Z}-\hat{\mathbf{ Z}}\right\Vert_{2}^{2}\left\Vert\vqsh\right\Vert_2^2\nonumber\\
    &\;\stackrel{(c)}{=}\left\Vert\mathbf{Z}-\hat{\mathbf{ Z}}\right\Vert_{2}^{2}\label{eq:eq12},
\end{align} where $(a)$ is obtained from the definition of $\mathbf{Z}$, and $(b)$ from matrix norm inequalities. As $\uA$ and $\vqsh$ are unitary, their 2-norms are unity (c).  
}

{{
Finally, $\left\Vert\mathbf{ Z}-\hat{\mathbf{ Z}}\right\Vert_{2}^{2}$ is bounded  as follows. Recall in Lemma~\ref{eq:lem2}, we established the lower bound on convergence rate $\alpha$ that ensures the strong convexity of $f(\mathbf{Z})$. This result, in turn, let us establish the error bound for $\left\Vert\mathbf{ Z}-\hat{\mathbf{ Z}}\right\Vert_{2}^{2}$, which provides the bound for sample complexity. We have 
\begin{align}
\frac{\alpha}{2} \left\Vert\mathbf{ Z}-\hat{\mathbf{ Z}}\right\Vert_{2}^{2}         
     & \stackrel{\text{(a)}}\leq
     {\left\Vert \mathbf{M}\mathbf{\Psi} - \uA\mathbf{Z}\vqsh^{\trasp}\mathbf{\Psi} \right\Vert}_{F}^{2} -{\left\Vert \mathbf{M}\mathbf{\Psi} - \uA\hat{\mathbf{Z}}\vqsh^{\trasp}\mathbf{\Psi} \right\Vert}_{F}^{2}\nonumber\\
     &\leq
     {\left\Vert \mathbf{M}\mathbf{\Psi} - \uA\mathbf{Z}\vqsh^{\trasp}\mathbf{\Psi} \right\Vert}_{F}^{2} \nonumber \\
     &\stackrel{(b)}{=} {\left\Vert \mathbf{M}\mathbf{\Psi} - \uA\uA^{\trasp}\mathbf{M}\vqsh\vqsh^{\trasp}\mathbf{\Psi} \right\Vert}_{F}^{2} \nonumber\\
     &={\left\Vert \mathbf{M}\mathbf{\Psi} - \mathbf{P}_{\uA}\mathbf{M}\mathbf{P}_{\vqsh}{\mathbf{\Psi}} \right\Vert}_{F}^{2} \nonumber\\
     &\leq {\left\Vert \mathbf{M} - \mathbf{P}_{\uA}\mathbf{M}\mathbf{P}_{\vqsh} \right\Vert}_{F}^{2} \nonumber\\
     &\stackrel{(c)}{\leq} 2\sigma^{2}_{r+1}\left(\mathbf{M}\right)\left(3+\frac{n}{d}\right)\nonumber\\
     &\;\;\;\;\;\;+32\sigma_{1}^{2}(\mathbf{M})\left\Vert\mathbf{E}\right\Vert_{F}^{2}\left(\frac{1}{\delta_1^2}+ \frac{\left\Vert\left(\mathbf{S\Psi}\right)^{\dagger}\mathbf{S}\right\Vert_{F}^{2}}{\sigma_{l}^2(\mathbf{QS})}\right)
\end{align} where (a) is from $f(\mathbf{Z}) - f(\hat{\mathbf{Z}}) \geq  \frac{\alpha}{2}{\left\Vert \mathbf{Z}-\hat{\mathbf{Z}} \right\Vert}^{2}_{2}$ in Definition~\ref{def:strong_convexity}. Since Gradient Descent reaches a stationary point, it follows that $\bigtriangledown f(\hat{\mathbf{Z}})=0$. And (b) is from our definition of $\mathbf{Z}$, $\mathbf{Z}=\uA^{\trasp}\mathbf{M}\vqsh$. Finally, (c) follows from Lemma~\ref{eq:lem1}.  

\begin{align}\label{eq:delta}
\Delta&\equiv2\sigma^{2}_{r+1}\left(\mathbf{M}\right)\left(3+\frac{n}{d}\right)\nonumber\\&\;\;\;\;\;\;\;+32\sigma_{1}^{2}(\mathbf{M})\left\Vert\mathbf{E}\right\Vert_{F}^{2}\left(\frac{1}{\delta_1^2}+ \frac{\left\Vert\left(\mathbf{S\Psi}\right)^{\dagger}\mathbf{S}\right\Vert_{F}^{2}}{\sigma_{l}^2(\mathbf{QS})}\right)
\end{align}
Then, we have,
\begin{align}
    \left\Vert\mathbf{ Z}-\hat{\mathbf{ Z}}\right\Vert_{F}^{2} \leq \frac{2\Delta}{\alpha}.\nonumber
\end{align}}}

{{
By plugging $\left\Vert\mathbf{ Z}-\hat{\mathbf{ Z}}\right\Vert_{F}^{2} \leq \frac{2\Delta}{\alpha}$ into \eqref{eq:eq12} and replacing $\Delta$ with \eqref{eq:delta}, we obtain the bound of $\left\Vert\mathbf{P}_{\uA}\mathbf{M}\mathbf{P}_{\vqsh}-\uA\hat{\mathbf{Z}}\vqsh^{\trasp}\right\Vert_{2}^{2}$ in Lemma~\ref{eq:lem3} as 
\begin{align}
&\Vert\mathbf{P}_{\mathbf{U}_{A}}\mathbf{ M}\mathbf{P}_{\vqsh}-\mathbf{U}_{A}\hat{\mathbf{ Z}}\vqsh^{\trasp}\Vert_{2}^{2}\nonumber\\ &\leq
\frac{4m}{d}\left(2\sigma^{2}_{r+1}\left(\mathbf{M}\right)\left(3+\frac{n}{d}\right)\right.\nonumber\\&\;\;\;\;\;\;\;\;\;\;\;+\left.32\sigma_{1}^{2}(\mathbf{M})\left\Vert\mathbf{E}\right\Vert_{F}^{2}\left(\frac{1}{\delta_1^2}+ \frac{\left\Vert\left(\mathbf{S\Psi}\right)^{\dagger}\mathbf{S}\right\Vert_{F}^{2}}{\sigma_{l}^2(\mathbf{QS})}\right)\right).
\end{align}

With $\alpha \geq \frac{d}{2m}$, We obtain Lemma~\ref{eq:lem3} provided that 
\begin{align}
    d \geq c_1{\hat{\mu}}^{2}r^{2}(\ln r), \nonumber
\end{align} and $c_1 \geq 0$.
}}

\bibliography{ref}

\begin{thebibliography}{10}
\providecommand{\url}[1]{#1}
\csname url@samestyle\endcsname
\providecommand{\newblock}{\relax}
\providecommand{\bibinfo}[2]{#2}
\providecommand{\BIBentrySTDinterwordspacing}{\spaceskip=0pt\relax}
\providecommand{\BIBentryALTinterwordstretchfactor}{4}
\providecommand{\BIBentryALTinterwordspacing}{\spaceskip=\fontdimen2\font plus
\BIBentryALTinterwordstretchfactor\fontdimen3\font minus
  \fontdimen4\font\relax}
\providecommand{\BIBforeignlanguage}[2]{{%
\expandafter\ifx\csname l@#1\endcsname\relax
\typeout{** WARNING: IEEEtran.bst: No hyphenation pattern has been}%
\typeout{** loaded for the language `#1'. Using the pattern for}%
\typeout{** the default language instead.}%
\else
\language=\csname l@#1\endcsname
\fi
#2}}
\providecommand{\BIBdecl}{\relax}
\BIBdecl

\bibitem{elmahdy2020matrix}
A.~Elmahdy, J.~Ahn, C.~Suh, and S.~Mohajer, ``Matrix completion with
  hierarchical graph side information,'' \emph{Advances in neural information
  processing systems}, vol.~33, pp. 9061--9074, 2020.

\bibitem{chiang2015matrix}
K.-Y. Chiang, C.-J. Hsieh, and I.~S. Dhillon, ``Matrix completion with noisy
  side information,'' \emph{Advances in neural information processing systems},
  vol.~28, 2015.

\bibitem{zhang2019inductive}
M.~Zhang and Y.~Chen, ``Inductive matrix completion based on graph neural
  networks,'' \emph{arXiv preprint arXiv:1904.12058}, 2019.

\bibitem{xu2013speedup}
M.~Xu, R.~Jin, and Z.-H. Zhou, ``Speedup matrix completion with side
  information: Application to multi-label learning,'' \emph{Advances in neural
  information processing systems}, vol.~26, 2013.

\bibitem{candes_exact_2009}
\BIBentryALTinterwordspacing
E.~J. Candès and B.~Recht, ``\BIBforeignlanguage{en}{Exact {Matrix}
  {Completion} via {Convex} {Optimization}},''
  \emph{\BIBforeignlanguage{en}{Foundations of Computational Mathematics}},
  vol.~9, no.~6, pp. 717--772, Dec. 2009. [Online]. Available:
  \url{http://link.springer.com/10.1007/s10208-009-9045-5}
\BIBentrySTDinterwordspacing

\bibitem{candes2010matrix}
E.~J. Candes and Y.~Plan, ``Matrix completion with noise,'' \emph{Proceedings
  of the IEEE}, vol.~98, no.~6, pp. 925--936, 2010.

\bibitem{cai2016structured}
T.~Cai, T.~T. Cai, and A.~Zhang, ``Structured matrix completion with
  applications to genomic data integration,'' \emph{Journal of the American
  Statistical Association}, vol. 111, no. 514, pp. 621--633, 2016.

\bibitem{farias2021learning}
V.~Farias, A.~Li, and T.~Peng, ``Learning treatment effects in panels with
  general intervention patterns,'' \emph{Advances in Neural Information
  Processing Systems}, vol.~34, pp. 14\,001--14\,013, 2021.

\bibitem{liu2017new}
G.~Liu, Q.~Liu, and X.~Yuan, ``A new theory for matrix completion,''
  \emph{Advances in Neural Information Processing Systems}, vol.~30, 2017.

\bibitem{quiton_matrix_2020}
\BIBentryALTinterwordspacing
S.~J. Quiton, U.~Mitra, and S.~Mallikarjun~Sharada, ``\BIBforeignlanguage{en}{A
  matrix completion algorithm to recover modes orthogonal to the minimum energy
  path in chemical reactions},'' \emph{\BIBforeignlanguage{en}{The Journal of
  Chemical Physics}}, vol. 153, no.~5, p. 054122, Aug. 2020. [Online].
  Available: \url{http://aip.scitation.org/doi/10.1063/5.0018326}
\BIBentrySTDinterwordspacing

\bibitem{bac2022matrix}
S.~Bac, S.~J. Quiton, K.~Kron, J.~Chae, U.~Mitra, and S.~M. Sharada, ``A matrix
  completion algorithm for efficient calculation of quantum and variational
  effects in chemical reactions,'' \emph{The Journal of Chemical Physics}, vol.
  156, no.~18, p. 184119, 2022.

\bibitem{quiton2022toward}
S.~J. Quiton, J.~Chae, S.~Bac, K.~Kron, U.~Mitra, and S.~M. Sharada, ``Toward
  efficient direct dynamics studies of chemical reactions: A novel matrix
  completion algorithm,'' \emph{Journal of Chemical Theory and Computation},
  2022.

\bibitem{natarajan2014inductive}
N.~Natarajan and I.~S. Dhillon, ``Inductive matrix completion for predicting
  gene--disease associations,'' \emph{Bioinformatics}, vol.~30, no.~12, pp.
  i60--i68, 2014.

\bibitem{foucart2020weighted}
S.~Foucart, D.~Needell, R.~Pathak, Y.~Plan, and M.~Wootters, ``Weighted matrix
  completion from non-random, non-uniform sampling patterns,'' \emph{IEEE
  Transactions on Information Theory}, vol.~67, no.~2, pp. 1264--1290, 2020.

\bibitem{negahban2012restricted}
S.~Negahban and M.~J. Wainwright, ``Restricted strong convexity and weighted
  matrix completion: Optimal bounds with noise,'' \emph{The Journal of Machine
  Learning Research}, vol.~13, no.~1, pp. 1665--1697, 2012.

\bibitem{xu2015cur}
M.~Xu, R.~Jin, and Z.-H. Zhou, ``Cur algorithm for partially observed
  matrices,'' in \emph{International Conference on Machine Learning}.\hskip 1em
  plus 0.5em minus 0.4em\relax PMLR, 2015, pp. 1412--1421.

\bibitem{matrix_recovery_sketch}
J.~A. Tropp, A.~Yurtsever, M.~Udell, and V.~Cevher, ``Practical sketching
  algorithms for low-rank matrix approximation,'' \emph{SIAM Journal on Matrix
  Analysis and Applications}, vol.~38, no.~4, pp. 1454--1485, 2017.

\bibitem{matrix_recovery_linear}
E.~J. Candes and Y.~Plan, ``Tight oracle inequalities for low-rank matrix
  recovery from a minimal number of noisy random measurements,'' \emph{IEEE
  Transactions on Information Theory}, vol.~57, no.~4, pp. 2342--2359, 2011.

\bibitem{drineas2008relative}
P.~Drineas, M.~W. Mahoney, and S.~Muthukrishnan, ``Relative-error cur matrix
  decompositions,'' \emph{SIAM Journal on Matrix Analysis and Applications},
  vol.~30, no.~2, pp. 844--881, 2008.

\bibitem{mahoney2009cur}
M.~W. Mahoney and P.~Drineas, ``Cur matrix decompositions for improved data
  analysis,'' \emph{Proceedings of the National Academy of Sciences}, vol. 106,
  no.~3, pp. 697--702, 2009.

\bibitem{garrett1980variational}
B.~C. Garrett and D.~G. Truhlar, ``Variational transition state theory. primary
  kinetic isotope effects for atom transfer reactions,'' \emph{Journal of the
  American Chemical Society}, vol. 102, no.~8, pp. 2559--2570, 1980.

\bibitem{gonzalez1991interpolated}
A.~Gonzalez-Lafont, T.~N. Truong, and D.~G. Truhlar, ``Interpolated variational
  transition-state theory: Practical methods for estimating variational
  transition-state properties and tunneling contributions to chemical reaction
  rates from electronic structure calculations,'' \emph{The Journal of Chemical
  Physics}, vol.~95, no.~12, pp. 8875--8894, 1991.

\bibitem{ongie_algebraic_2017}
\BIBentryALTinterwordspacing
G.~Ongie, ``Algebraic {Variety} {Models} for {High}-{Rank} {Matrix}
  {Completion} {MATLAB} code,'' Jul. 2017, (accessed 2021-04-30). [Online].
  Available: \url{https://github.com/gregongie/vmc}
\BIBentrySTDinterwordspacing

\bibitem{miller1980reaction}
W.~H. Miller, N.~C. Handy, and J.~E. Adams, ``Reaction path hamiltonian for
  polyatomic molecules,'' \emph{The Journal of chemical physics}, vol.~72,
  no.~1, pp. 99--112, 1980.

\bibitem{page1988evaluating}
M.~Page and J.~W. McIver~Jr, ``On evaluating the reaction path hamiltonian,''
  \emph{The Journal of chemical physics}, vol.~88, no.~2, pp. 922--935, 1988.

\bibitem{boyd2004convex}
S.~Boyd and L.~Vandenberghe, \emph{Convex optimization}.\hskip 1em plus 0.5em
  minus 0.4em\relax Cambridge university press, 2004.

\bibitem{tropp2011improved}
J.~A. Tropp, ``Improved analysis of the subsampled randomized hadamard
  transform,'' \emph{Advances in Adaptive Data Analysis}, vol.~3, no. 01n02,
  pp. 115--126, 2011.

\bibitem{stewart1998perturbation}
G.~W. Stewart, ``Perturbation theory for the singular value decomposition,''
  Tech. Rep., 1998.

\bibitem{eckart1936approximation}
C.~Eckart and G.~Young, ``The approximation of one matrix by another of lower
  rank,'' \emph{Psychometrika}, vol.~1, no.~3, pp. 211--218, 1936.

\bibitem{cs_covariance}
M.~Azizyan, A.~Krishnamurthy, and A.~Singh, ``Extreme compressive sampling for
  covariance estimation,'' \emph{IEEE Transactions on Information Theory},
  vol.~64, no.~12, pp. 7613--7635, 2018.

\bibitem{brand2006fast}
M.~Brand, ``Fast low-rank modifications of the thin singular value
  decomposition,'' \emph{Linear algebra and its applications}, vol. 415, no.~1,
  pp. 20--30, 2006.

\bibitem{xu2015cur_sup}
M.~Xu, R.~Jin, and Z.-H. Zhou, ``Supplementary of cur algorithm for partially
  observed matrices,'' in \emph{International Conference on Machine
  Learning}.\hskip 1em plus 0.5em minus 0.4em\relax PMLR, 2015, pp. 1412--1421.

\bibitem{stewart1990matrix}
G.~W. Stewart, ``Matrix perturbation theory,'' 1990.

\bibitem{Rencher2012methods}
C.~Rencher, Alvin and William.F, ``Methods of multivariate analysis,'' 2012.

\bibitem{negahban2012unified}
S.~N. Negahban, P.~Ravikumar, M.~J. Wainwright, and B.~Yu, ``A unified
  framework for high-dimensional analysis of $ m $-estimators with decomposable
  regularizers,'' \emph{Statistical science}, vol.~27, no.~4, pp. 538--557,
  2012.

\bibitem{Roger1991matrix}
R.~A. Horn and C.~R. Johnson, ``Topics in matrix analysis,'' 1991.

\end{thebibliography}

\end{document}